\documentclass[aps,jmp,superscriptaddress,longbibliography]{revtex4-2}
\usepackage{amscd}
\usepackage{amssymb}
\usepackage{amsmath}
\usepackage{bm}
\usepackage[margin=1in]{geometry}
\allowdisplaybreaks
\usepackage[T1]{fontenc}
\usepackage[latin1]{inputenc}
\usepackage{graphicx}
\usepackage{graphics}
\usepackage{color}
\usepackage{braket}
\usepackage{enumitem}
\usepackage{ifpdf}
\ifpdf
\else
    
\fi

\usepackage{tikz}

\ifpdf
    \usepackage[hypertexnames=false]{hyperref}
    \hypersetup{
        colorlinks=true,
        linkcolor=blue,
        citecolor=blue,
        urlcolor=blue
    }
\fi
\newcommand{\erfc}{\operatorname{erfc}}
\def\be{\begin{equation}}
\def\ee{\end{equation}}
\def\ba{\begin{eqnarray}}
\def\ea{\end{eqnarray}}
\def\>{\rangle}
\def\<{\langle}
\def\n{\nonumber}

\begin{document}

\title{Exact Ionization Amplitudes for a Delta-Function Well in an Arbitrary-Strength DC Electric Field}

\author{Ilki Kim}
\email{hannibal.ikim@gmail.com} \affiliation{Department of Physics,
North Carolina A$\&$T State University, Greensboro, NC 27411}
\author{Gerald J. Iafrate}
\affiliation{\mbox{Department of Electrical and Computer
Engineering, North Carolina State University, Raleigh, NC 27695}}

\date{\today}

\begin{abstract}
We investigate finite-time field-induced ionization from a
one-dimensional attractive delta-function well subjected to an
arbitrary-strength uniform dc electric field. Our objective is to
determine the physical bound-state survival amplitude \(a_b(t)\) at
an arbitrary observation time without first constructing the
complete time-dependent propagator or wavefunction. Using the
gauge-equivalent Kramers--Henneberger representation, in which the
field is incorporated into the motion of the contact point, we
develop an exact amplitude-centered formulation of the problem. The
moving-contact dynamics is reduced to a closed Volterra equation,
and exact endpoint-phase factorization organizes its complete
chronological rescattering history into an ordinary relative-time
convolution hierarchy and an exact resolvent. Introducing the
accumulated bound-state amplitude permits its two-time domain to be
reorganized in terms of the relative time and its complementary
center time; using these variables then reduces the final contact
contribution to an explicit boundary integral. All spatial
integrations are performed analytically: the field-driven,
contact-free contribution is obtained in closed form in terms of the
Faddeeva function, while the direct and repeated-rescattering
contributions are given by explicitly evaluable time integrals. The
result is valid for arbitrary dc-field strength and arbitrary finite
observation time, without a weak-field expansion, rescattering
truncation, or asymptotic-time approximation, and is verified by the
exact field-free limit and an independent numerical solution of the
original physical Volterra integral equation. This formulation
provides an exact nonperturbative solution of a fundamental
ionization model and demonstrates how a suitable representation and
temporal organization can expose otherwise hidden analytical
structure in driven quantum dynamics.
\end{abstract}

\maketitle

\section{Introduction}
\label{sec:intro}
%
Field-induced ionization of bound quantum states is one of the
canonical problems of nonperturbative quantum dynamics. As a
fundamental manifestation of the interaction between matter and
electromagnetic fields, it occupies a central position in atomic,
molecular, optical, and strong-field physics. Unlike transitions
between discrete bound states, ionization transfers probability from
an initially localized state into a continuum of scattering states
and therefore combines coherent quantum evolution with the dynamical
role of an unbounded spectrum. Despite decades of intensive study,
obtaining an explicit analytical description of this process at
arbitrary finite times remains a challenging problem.

The modern theoretical description of strong-field ionization has
developed through several complementary approaches. Beginning with
the pioneering work of Keldysh, subsequent formulations by
Perelomov, Popov, and Terent'ev (PPT) and by Ammosov, Delone, and
Krainov (ADK), together with the strong-field approaches developed
independently by Faisal and Reiss, established the foundations for
the description of tunneling, multiphoton, and nonperturbative
ionization phenomena \cite{KEL65,PPT66,ADK86,FAI73,REI80}. The
approaches associated with Keldysh, Faisal, and Reiss are commonly
grouped under the Keldysh--Faisal--Reiss (KFR) framework.
Complementary studies of atoms in intense laser fields by Gavrila
and collaborators further advanced the understanding of atomic
stabilization, high-frequency laser--matter interaction, and
nonperturbative strong-field dynamics \cite{GAV92,GAV02}. Together,
these developments provide much of the theoretical foundation for
contemporary strong-field ionization theory.

Considerable progress has consequently been achieved in the
description of ionization rates, tunneling probabilities,
quasienergy spectra, Floquet states, semiclassical trajectories, and
asymptotic transition amplitudes. A different analytical challenge
arises, however, when the objective is the explicit finite-time
evolution of the initially occupied bound state. The corresponding
bound-state survival amplitude \(a_b(t)\) directly measures the
coherent amplitude remaining in that state and determines the
ionization probability through
\(\mathcal{P}_{\mbox{\scriptsize{ion}}}(t) = 1 - |a_b(t)|^2\). Its
finite-time evaluation must retain the coherent bound-to-continuum
dynamics, the memory associated with repeated interactions with the
binding potential, and the rapidly oscillatory phases generated by
the external field. These features make an explicit finite-time
amplitude substantially more difficult to obtain analytically than
many asymptotic quantities.

Part of this difficulty arises from the conventional analytical
organization of the problem. Exact formulations commonly proceed by
constructing the complete time-dependent propagator or wavefunction
and only afterward projecting the resulting dynamics onto the bound
state of interest. This route is formally natural and contains the
complete dynamical information, but precisely for that reason it
requires the determination of considerably more information than is
needed for a single projected amplitude. In strongly driven systems,
where chronological propagation generates highly oscillatory
two-time phase structures and repeated rescattering, the
construction of this more general object can become the dominant
analytical difficulty even when the desired observable is only the
bound-state survival amplitude.

The present work therefore adopts a different analytical
organization. Instead of taking the complete chronological
propagator as the primary object and extracting \(a_b(t)\) only at
the end, we ask whether the physical bound-state survival amplitude
can be organized and evaluated directly. As shown below, this
amplitude-centered formulation becomes particularly effective in the
Kramers--Henneberger representation, where the external field is
transferred to the motion of the binding potential. By introducing
the accumulated bound-state amplitude and reorganizing its two-time
structure in terms of relative and center times, the chronological
rescattering dynamics can be reduced exactly to an ordinary
relative-time convolution hierarchy. The physical amplitude can
thereby be obtained without first constructing the complete
time-dependent propagator.

To examine these questions concretely, we consider a one-dimensional
attractive delta-function well in the presence of an
arbitrary-strength uniform dc electric field. Despite its
mathematical simplicity, the model retains the essential ingredients
of field-induced ionization while preserving the principal
analytical challenges of finite-time dynamics. In the absence of an
external field, it supports a single localized bound state; when the
dc field is applied, this state becomes coupled to the continuum,
providing a minimal nontrivial realization of quantum ionization
dynamics. Accordingly, the delta-function well serves not only as
the specific physical system considered here, but also as an
analytically complete testbed in which more general strategies for
finite-time driven dynamics can be developed and examined.

Over the past several decades, this model has served as a valuable
laboratory for investigating field-induced decay, tunneling,
resonant transport, and exactly solvable time-dependent quantum
problems. Beginning with the pioneering work of Geltman
\cite{GEL77,GEL78}, numerous exact and semi-exact formulations have
been developed using propagators, Green functions, Volterra integral
equations, Laplace-transform techniques, and numerical inversion
procedures \cite{LUD87,ELK88,SUS90,KLE94,ENG95,ROK00}. These studies
have demonstrated that remarkably rich ionization dynamics can
emerge even from the simplest short-range potential. At the same
time, they illustrate a persistent analytical difficulty: formally
exact dynamical equations can often be established, whereas an
explicit finite-time expression for the physical bound-state
survival amplitude remains considerably more difficult to obtain.

One source of this difficulty is methodological rather than
physical. In many such formulations, the complete chronological
propagator or the full time-dependent wavefunction is constructed
first, and the desired bound-state amplitude is obtained only
afterward by projection. Although formally exact, this procedure
requires solving a substantially richer dynamical problem than is
necessary for determining a single projected physical amplitude.

In contrast to the well-known scalar- and vector-potential
formulations, the Kramers--Henneberger (KH) representation transfers
the influence of the external field to a time-dependent displacement
of the binding potential \cite{HEN68,BRE92}. For the present
delta-function well, this transformation converts the field-driven
problem into one involving a moving contact. Between successive
contact interactions the propagation is therefore free, with no
explicit field driving, while the influence of the external field is
retained through the trajectory of the moving contact. This
representation has long provided an elegant framework for
strong-field dynamics, and its continuing relevance is reflected in
recent renewed interest in KH-based formulations of nonperturbative
laser--matter interaction \cite{ARG25}. This moving-contact
structure provides the starting point for the amplitude-centered
formulation developed below.

We therefore retain the KH representation while departing from the
conventional propagator-centered route. Rather than constructing the
complete chronological propagator and subsequently extracting the
observable of interest, we formulate the problem directly in terms
of the physical bound-state survival amplitude. As in other
gauge-equivalent representations, the dynamics can first be reduced
to a closed Volterra equation for the wavefunction at the
interaction point, whose solution enters directly into the
bound-state projection. The distinctive role of the KH
representation emerges at the next stage. Introducing the
accumulated bound-state amplitude enlarges the temporal description
from a single observation time to an exact two-time domain, which
can be reorganized in terms of the relative and center times. In the
KH representation, this reorganization renders the field-dependent
phase of each propagation interval linear in its relative time when
its corresponding center time is held fixed.

The decisive simplification occurs when the physical Volterra
equation is expanded into its chronological rescattering series and
inserted directly into the bound-state projection. Although
successive propagation intervals generally carry different center
times, the dependence on the intermediate center times cancels
exactly when the complete chronological history is assembled. The
remaining field-dependent phases depend only on the successive
relative-time intervals, allowing repeated contact interactions to
be organized as an ordinary convolution hierarchy generated by a
single relative-time kernel. The resulting all-orders contact
contribution can consequently be expressed through fully specified,
analytically evaluable convolution kernels and an exact convolution
resolvent, without first constructing the complete chronological
propagator.

The resulting formulation provides an exact finite-time
representation of the physical bound-state survival amplitude,
consisting of an explicit contact-free contribution and an
all-orders contact contribution that separates naturally into direct
and repeated-rescattering terms. No weak-field expansion,
rescattering truncation, or asymptotic-time approximation is
required. The significance of the delta-function model in the
present context is therefore not only that it permits an exact
analytical treatment, but that it provides an analytically complete
setting in which the amplitude-centered temporal reorganization can
be established explicitly. This strategy may provide a useful
foundation for more general formulations of nonperturbative driven
quantum dynamics.

The remainder of this paper is organized as follows.
Sec.~\ref{sec:kh_formulation} reviews the gauge formulations,
introduces the Kramers--Henneberger representation, and derives the
exact moving-contact Volterra integral equation governing the
contact dynamics. Sec.~\ref{sec:uw} develops the
accumulated-amplitude formulation and the exact reorganization into
relative- and center-time variables. Sec.~\ref{sec:exact_solution}
derives the all-orders convolution hierarchy, obtains the resulting
exact finite-time bound-state amplitude, and examines its physical
and computational consequences. Finally, Sec.~\ref{sec:conclusions}
summarizes the principal results and concludes the paper.

\section{Gauge Formulations and the Kramers--Henneberger
Representation} \label{sec:kh_formulation}
%
\subsection{Scalar-potential gauge and laboratory survival amplitude}
\label{subsec:model_survival}
We consider a particle of mass \(m\) bound by a one-dimensional
attractive delta-function potential and subjected to an
arbitrary-strength uniform dc electric field. In the
scalar-potential gauge, the Hamiltonian is
\begin{equation}
    \hat{H}_{\mbox{\scriptsize{s}}}(t)
    =
    \frac{\hat{p}^2}{2 m}
    -
    g \delta(\hat{x})
    -
    \hat{x} F(t)\,,
    \label{eq:scalar_hamiltonian}
\end{equation}
where \(g > 0\) denotes the strength of the attractive
delta-function potential, and
\begin{equation}
    F(t)
    =
    F_0\,\Theta(t)
    \label{eq:dc_field}
\end{equation}
is the externally applied uniform dc electric field. The field is
switched on at \(t=0\) and remains constant thereafter.

The corresponding field-free Hamiltonian is
\begin{equation}
    \hat{H}_b
    =
    \frac{\hat{p}^2}{2 m}
    -
    g \delta(\hat{x})\,,
    \label{eq:field_free_hamiltonian}
\end{equation}
which supports a single normalized bound state
\begin{equation}
    \psi_b(x)
    =
    \sqrt{\kappa}\,
    e^{-\kappa |x|}\,.
    \label{eq:bound_state}
\end{equation}
Here
\begin{equation}
    \kappa
    =
    \frac{m g}{\hbar^2}\,,
    \label{eq:kappa_definition}
\end{equation}
and the corresponding bound-state energy is
\begin{equation}
    E_b
    =
    -\frac{\hbar^2\kappa^2}{2m}\,.
    \label{eq:bound_state_energy}
\end{equation}
A derivation of the field-free bound state is provided in
Appendix~\ref{app:field_free_case}. In addition to this discrete
bound state, this field-free Hamiltonian possesses a continuum of
positive-energy scattering states. The bound and continuum
eigenstates together form a complete set, as discussed in detail,
e.g., in Refs.~\cite{DAM75,PAT00,GOT03}. This spectral structure
provides the natural basis for the field-induced coupling between
the bound and continuum sectors that underlies the ionization
dynamics considered below.

The principal physical quantity of interest is the laboratory-frame
bound-state survival amplitude
\begin{equation}
    a_b(t)
    =
    \langle\psi_b|
    \hat{U}_{\mbox{\scriptsize{s}}}(t,0)
    |\psi_b\rangle
    =
    \langle\psi_b|
    \psi_{\mbox{\scriptsize{s}}}(t)
    \rangle\,,
    \label{eq:physical_survival_amplitude}
\end{equation}
where \(\hat{U}_{\mbox{\scriptsize{s}}}(t,0)\) is the scalar-gauge
time-evolution operator, and
$|\psi_{\mbox{\scriptsize{s}}}(t)\rangle =
\hat{U}_{\mbox{\scriptsize{s}}}(t,0)|\psi(0)\rangle$ with
$|\psi(0)\rangle = |\psi_b\rangle$.

\subsection{Comparison with vector-potential gauge}
\label{subsec:vector_gauge}
The same field-driven dynamics can equally be described in the
vector-potential gauge by the Hamiltonian
\begin{equation}
    \hat{H}_{\mbox{\scriptsize{v}}}(t)
    =
    \frac{\left[\hat{p} + p_c(t)\right]^2}{2 m}
    -
    g \delta(\hat{x})\,,
    \label{eq:vector_hamiltonian}
\end{equation}
where the field-induced classical momentum is
\begin{equation}
    p_c(t)
    =
    \int_0^t F(\tau)\,d\tau\,.
    \label{eq:pc_definition}
\end{equation}
This \(c\)-number momentum shift corresponds to the spatially
uniform vector potential through
\begin{equation}
    p_c(t)
    =
    -\frac{q}{c} A(t)\,.
    \label{eq:pc_vector_potential}
\end{equation}
For the dc field defined in Eq.~\eqref{eq:dc_field}, we have
\begin{subequations}
\begin{equation}
    p_c(t)
    =
    F_0 t\,.
    \label{eq:pc_dc}
\end{equation}
The resulting position shift is
\begin{equation}
    x_c(t)
    =
    \frac{1}{m}
    \int_0^t p_c(\tau)\,d\tau
    =
    \frac{F_0 t^2}{2 m}\,,
    \label{eq:xc_dc}
\end{equation}
and the associated classical action is
\begin{equation}
    s_c(t)
    =
    \frac{1}{2 m}
    \int_0^t p_c^2(\tau)\,d\tau
    =
    \frac{F_0^2 t^3}{6 m}\,.
    \label{eq:action_dc}
\end{equation}
\end{subequations}

Starting from the initial bound state, the vector-gauge state
evolves as
\begin{equation}
    |\psi_{\mbox{\scriptsize{v}}}(t)\rangle
    =
    \hat{U}_{\mbox{\scriptsize{v}}}(t,0)
    |\psi_b\rangle\,.
    \label{eq:vector_evolved_state}
\end{equation}
The scalar- and vector-gauge state vectors are related by the
time-dependent unitary transformation
\begin{subequations}
\begin{equation}
    |\psi_{\mbox{\scriptsize{s}}}(t)\rangle
    =
    \hat{U}_{{\mbox{\scriptsize{s}}}
    \leftarrow{\mbox{\scriptsize{v}}}}(t)
    |\psi_{\mbox{\scriptsize{v}}}(t)\rangle
    \;\;\; ;\;\;\;
    \hat{U}_{{\mbox{\scriptsize{s}}}
    \leftarrow{\mbox{\scriptsize{v}}}}(t)
    =
    \exp\!\left[
        \frac{i}{\hbar}p_c(t)\hat{x}
    \right]\,.
    \label{eq:scalar_vector_state_relation}
\end{equation}
In the coordinate representation,
\begin{equation}
    \psi_{\mbox{\scriptsize{s}}}(x,t)
    =
    \exp\!\left[
        \frac{i}{\hbar}p_c(t)x
    \right]
    \psi_{\mbox{\scriptsize{v}}}(x,t)\,.
    \label{eq:scalar_vector_wavefunction_relation}
\end{equation}
\end{subequations}
For a detailed discussion of scalar- versus vector-gauge problems in
quantum optics, see Ref.~\cite{SCH84}.

The laboratory survival amplitude therefore becomes
\begin{equation}
    a_b(t)
    =
    \langle\psi_b|
    \hat{U}_{{\mbox{\scriptsize{s}}}
    \leftarrow{\mbox{\scriptsize{v}}}}(t)
    \hat{U}_{\mbox{\scriptsize{v}}}(t,0)
    |\psi_b\rangle
    =
    \langle\psi_b|
    \hat{U}_{{\mbox{\scriptsize{s}}}
    \leftarrow{\mbox{\scriptsize{v}}}}(t)
    |\psi_{\mbox{\scriptsize{v}}}(t)\rangle\,.
    \label{eq:survival_amplitude_vector_gauge}
\end{equation}
This gauge-consistent expression has a direct physical meaning.
After the dc field is switched off at the observation time, the
evolved state is projected onto the field-free bound state
\(\psi_b\), and the remaining bound-state probability is
\(|a_b(t)|^2\). The corresponding ionization probability is
therefore \(\mathcal{P}_{\mbox{\scriptsize{ion}}}(t) = 1 -
|a_b(t)|^2\).

Accordingly, the projection
\(\langle\psi_b|\psi_{\mbox{\scriptsize{v}}}(t)\rangle\) is not, by
itself, the physical survival amplitude. Indeed, the formally
related vector-gauge quantity
\begin{equation}
    \mathcal{P}_{\mbox{\scriptsize{v}}}(t)
    =
    1 -
    \left|
        \langle\psi_b|
        \psi_{\mbox{\scriptsize{v}}}(t)\rangle
    \right|^2
    \label{eq:vector_gauge_probability}
\end{equation}
does not represent the same physical measurement. As pointed out in
\cite{ELB87}, it instead corresponds to a different final operation,
namely a formal turning off of the temporally accumulated vector
potential rather than the physical electric field. The physical
survival amplitude is therefore obtained from the gauge-consistent
matrix element in Eq.~\eqref{eq:survival_amplitude_vector_gauge},
rather than from the naive projection of a gauge-transformed
wavefunction onto the untransformed bound state. Thus, the physical
quantity is the same in all gauge-equivalent representations,
provided that the final projection is transformed consistently. This
distinction will remain important when the same physical amplitude
is expressed in the Kramers--Henneberger representation.

The vector-potential gauge nevertheless provides a convenient route
to the field-only propagator. In momentum space, the corresponding
evolution factor is
\begin{equation}
    U_{\mbox{\scriptsize{v}},\scriptstyle{F}}(p;t,\tau)
    =
    \exp\!\left[
        -\frac{i}{\hbar}
        \int_{\tau}^{t}
        \frac{\left[p + p_c(t')\right]^2}{2 m}\,dt'
    \right]\,.
    \label{eq:vector_field_evolution_factor}
\end{equation}
Its Fourier transformation then gives
\begin{align}
    K_{\mbox{\scriptsize{v}},\scriptstyle{F}}(x,t|x',\tau)
    &=
    \frac{1}{2 \pi \hbar}
    \int_{-\infty}^{\infty} dp\,
    \exp\!\left[
        \frac{i}{\hbar} p (x - x')
    \right]\,
    U_{\mbox{\scriptsize{v}},\scriptstyle{F}}(p;t,\tau)
    \nonumber\\
    &=
    \sqrt{\frac{m}{2 \pi i \hbar (t - \tau)}}\,
    \exp\!\left[
        -\frac{i}{\hbar}
        \left\{
            s_c(t) - s_c(\tau)
        \right\}
    \right]\,
    \exp\!\left[
        \frac{i}{\hbar}
        \frac{m}{2 (t - \tau)}
        \left(
            [x - x_c(t)]
            -
            [x' - x_c(\tau)]
        \right)^2
    \right]\,.
    \label{eq:vector_field_propagator}
\end{align}
Gauge covariance with
Eq.~\eqref{eq:scalar_vector_wavefunction_relation} yields the
scalar-gauge field-only propagator
\begin{equation}
    K_{\mbox{\scriptsize{s}},\scriptstyle{F}}(x,t|x',\tau)
    =
    \exp\!\left[
        \frac{i}{\hbar} x p_c(t)
    \right]\,
    K_{\mbox{\scriptsize{v}},\scriptstyle{F}}(x,t|x',\tau)\,
    \exp\!\left[
        -\frac{i}{\hbar}x' p_c(\tau)
    \right]\,.
    \label{eq:scalar_from_vector_propagator}
\end{equation}
Using Eqs.~\eqref{eq:pc_dc}--\eqref{eq:action_dc}, this expression
reduces, for the dc field, to
\begin{align}
    K_{\mbox{\scriptsize{s}},F}(x,t|x',\tau)
    &=
    \sqrt{
        \frac{m}{2 \pi i \hbar (t - \tau)}
    }\,
    \exp\!\left[
        \frac{i m (x - x')^2}{2 \hbar (t - \tau)}
    \right]\,
    \exp\!\left[
        -\frac{i}{\hbar}
        \left\{
            \frac{F_0^2 (t - \tau)^3}{24 m}
            -
            \frac{F_0 (x + x')}{2}(t - \tau)
        \right\}
    \right]\,.
    \label{eq:scalar_dc_field_propagator}
\end{align}
In particular, propagation from the contact point \(x' = 0\) is
described by
\begin{equation}
    K_{\mbox{\scriptsize{s}},\scriptstyle{F}}(x,t|0,\tau)
    =
    K_0(x,t|0,\tau)\,
    \exp\!\left[
        \frac{i}{\hbar}
        \left\{
            \frac{F_0 x}{2}(t - \tau)
            -
            \frac{F_0^2}{24 m}(t - \tau)^3
        \right\}
    \right]\,,
    \label{eq:scalar_contact_propagator}
 \end{equation}
where
\begin{equation}
    K_0(x,t|0,\tau)
    =
    \sqrt{
        \frac{m}{2 \pi i \hbar (t - \tau)}
    }\,
    \exp\!\left[
        \frac{i m x^2}{2 \hbar (t - \tau)}
    \right]
    \label{eq:free_contact_propagator}
\end{equation}
is the one-dimensional free propagator evaluated at \(x' = 0\).

The exact scalar-gauge propagator satisfies the Lippmann--Schwinger
integral equation \cite{KLE94}
\begin{equation}
\label{eq:general_scalar_lippmann_schwinger_propagator}
    \mathcal{K}_{\mbox{\scriptsize{s}}}(x,t|x',0)
    =
    K_{\mbox{\scriptsize{s}},\scriptstyle{F}}(x,t|x',0)
    -
    \frac{i}{\hbar}
    \int_0^t d\tau
    \int_{-\infty}^{\infty} dx''\,
    K_{\mbox{\scriptsize{s}},\scriptstyle{F}}(x,t|x'',\tau)\,
    V(x'')\,
    \mathcal{K}_{\mbox{\scriptsize{s}}}(x'',\tau|x',0)\,,
\end{equation}
where \(V(x)\) denotes the binding potential. For the delta-function
well, \(V(x) = -g \delta(x)\),
Eq.~\eqref{eq:general_scalar_lippmann_schwinger_propagator} reduces
to \cite{ELK88}
\begin{equation}
    \mathcal{K}_{\mbox{\scriptsize{s}}}(x,t|x',0)
    =
    K_{\mbox{\scriptsize{s}},\scriptstyle{F}}(x,t|x',0)
    +
    \frac{i g}{\hbar}
    \int_0^t d\tau\,
    K_{\mbox{\scriptsize{s}},\scriptstyle{F}}(x,t|0,\tau)\,
    \mathcal{K}_{\mbox{\scriptsize{s}}}(0,\tau|x',0)\,.
    \label{eq:scalar_lippmann_schwinger_propagator}
\end{equation}
The corresponding scalar-gauge wavefunction is then given by
\begin{equation}
    \psi_{\mbox{\scriptsize{s}}}(x,t)
    =
    \int_{-\infty}^{\infty} dx'\,
    \mathcal{K}_{\mbox{\scriptsize{s}}}(x,t|x',0)\,
    \psi_b(x')\,,
    \label{eq:scalar_wavefunction_from_full_propagator}
\end{equation}
where the homogeneous contribution is
\begin{equation}
    \phi_{\mbox{\scriptsize{s}}}(x,t)
    =
    \int_{-\infty}^{\infty} dx'\,
    K_{\mbox{\scriptsize{s}},\scriptstyle{F}}(x,t|x',0)\,
    \psi_b(x')\,.
    \label{eq:scalar_homo_wavefunction_from_full_propagator}
\end{equation}
Eq.~\eqref{eq:scalar_lippmann_schwinger_propagator} is formally
exact, but it also illustrates the central analytical difficulty of
the conventional propagator-first formulation. To appreciate the
origin of this difficulty, it is useful first to recall the
field-free case. There, the corresponding kernel depends only on the
time difference \((t - \tau)\), allowing direct application of the
Laplace convolution theorem and yielding an exact time-domain
solution through the inverse Laplace transform \cite{ELB88}. By
contrast, for \(F_0 \neq 0\), the contact kernel in
Eq.~\eqref{eq:scalar_contact_propagator} acquires the highly
oscillatory cubic phase
\begin{equation}
    \exp\!\left[
        -\frac{i}{\hbar}
        \frac{F_0^2}{24 m} (t - \tau)^3
    \right]\,.
    \label{eq:scalar_cubic_phase}
\end{equation}
Although this factor remains a function only of the relative time
\((t - \tau)\) and therefore admits an ordinary Laplace transform,
the transform is non-elementary. It can be expressed exactly in
terms of generalized hypergeometric functions, while its termwise
expansion in powers of the cubic-phase parameter yields only a
formal asymptotic series rather than a convergent power series.
Thus, the Laplace convolution theorem remains applicable, but the
transformed kernel and its inverse do not by themselves provide a
practically simple analytical construction of the complete
propagator; see also the discussion following
Eq.~\eqref{eq:kF_laplace}.

Next, evaluating Eq.~\eqref{eq:scalar_lippmann_schwinger_propagator}
at the contact position \(x = 0\) gives the closed Volterra integral
equation
\begin{equation}
    \mathcal{K}_{\mbox{\scriptsize{s}}}(0,t|x',0)
    =
    K_{\mbox{\scriptsize{s}},\scriptstyle{F}}(0,t|x',0)
    +
    \frac{i g}{\hbar}
    \int_0^t d\tau\,
    K_{\mbox{\scriptsize{s}},\scriptstyle{F}}(0,t|0,\tau)\,
    \mathcal{K}_{\mbox{\scriptsize{s}}}(0,\tau|x',0)\,.
    \label{eq:scalar_contact_equation}
\end{equation}
Once \(\mathcal{K}_{\mbox{\scriptsize{s}}}(0,t|x',0)\) is determined
from Eq.~\eqref{eq:scalar_contact_equation}, its substitution into
Eq.~\eqref{eq:scalar_lippmann_schwinger_propagator} yields the
complete propagator
\(\mathcal{K}_{\mbox{\scriptsize{s}}}(x,t|x',0)\). However, because
the field propagator
\(K_{\mbox{\scriptsize{s}},\scriptstyle{F}}(0,t|0,\tau)\) contains
the highly oscillatory cubic phase in
Eq.~\eqref{eq:scalar_cubic_phase}, the contact equation
\eqref{eq:scalar_contact_equation} does not by itself provide a
practically simple analytical construction of the complete
propagator. This difficulty is further amplified by the underlying
physics of repeated interactions with the delta-function potential.
The exact propagator contains an infinite hierarchy of highly
oscillatory multiple-scattering histories, including repeated
bound--continuum transitions and continuum--continuum rescattering
processes. A detailed momentum-space representation of this
hierarchy is given in Appendix~\ref{app:momentum_propagator}.
Consequently, although the conventional formulation is formally
exact, the explicit propagator itself becomes the principal
analytical bottleneck: infinitely many ionization--recombination
histories must be organized before the physical bound-state
amplitude can be extracted. Within this propagator-level
organization, the simpler relative-time structure relevant to the
projected amplitude remains obscured by the nested chronological
integrations. The issue is therefore not the absence of a formal
transform or an exact integral equation, but rather the analytical
organization inherent in the propagator-first approach. This
observation motivates the moving-contact reformulation in the
Kramers--Henneberger representation, in which the uniform dc field
is absorbed into the motion of the contact point, as presented in
Subsecs.~\ref{subsec:kh_representation}--\ref{subsec:remaining_structure}.
This reformulation preserves the exact dynamics while providing the
appropriate starting point for the amplitude-centered temporal
reorganization developed in Sec.~\ref{sec:uw}.

\subsection{Kramers--Henneberger representation}
\label{subsec:kh_representation}
We now transform the vector-gauge formulation to the
Kramers--Henneberger (KH) representation. Let
\(\hat{U}_{\mbox{\scriptsize
KH}\leftarrow{\mbox{\scriptsize{v}}}}(t)\) denote the unitary
operator that maps the vector-gauge state to the KH state:
\begin{equation}
    |\psi_{\mbox{\scriptsize KH}}(t) \rangle
    =
    \hat{U}_{\mbox{\scriptsize KH}\leftarrow{\mbox{\scriptsize{v}}}}(t)\,
    |\psi_{\mbox{\scriptsize{v}}}(t)\rangle\,.
    \label{eq:kh_state_definition}
\end{equation}
Substituting $|\psi_{\mbox{\scriptsize{v}}}(t)\rangle$ in Eq.
\eqref{eq:kh_state_definition} into the vector-gauge Schr\"{o}dinger
equation, it follows that
\begin{equation}
    \hat{H}_{\mbox{\scriptsize{KH}}}(t)
    =
    \hat{U}_{\mbox{\scriptsize{KH}}\leftarrow{\mbox{\scriptsize{v}}}}(t)\,
    \hat{H}_{\mbox{\scriptsize{v}}}(t)\,
    \hat{U}_{\mbox{\scriptsize{KH}}\leftarrow{\mbox{\scriptsize{v}}}}^{\dagger}(t)
    +
    i \hbar\,
    \dot{\hat{U}}_{\mbox{\scriptsize{KH}}\leftarrow{\mbox{\scriptsize{v}}}}(t)\,
    \hat{U}_{\mbox{\scriptsize{KH}}\leftarrow{\mbox{\scriptsize{v}}}}^{\dagger}(t)\,.
    \label{eq:kh_hamiltonian_general}
\end{equation}

We choose
\begin{equation}
    \hat{U}_{\mbox{\scriptsize{KH}}\leftarrow{\mbox{\scriptsize{v}}}}(t)
    =
    \exp\!\left[
        \frac{i}{\hbar} \hat{p}\,x_c(t)
    \right]\,
    \exp\!\left[
        \frac{i}{\hbar} s_c(t)
    \right]\,,
    \label{eq:kh_vector_transformation}
\end{equation}
where \(x_c(t)\) and \(s_c(t)\) are given in Eqs.~\eqref{eq:xc_dc}
and \eqref{eq:action_dc}, respectively. The first factor translates
the coordinate operator according to
\begin{equation}
    e^{\frac{i}{\hbar} \hat{p} x_c(t)}\,
    \hat{x}\,
    e^{-\frac{i}{\hbar} \hat{p} x_c(t)}
    =
    \hat{x} + x_c(t)\;\;\; ;\;\;\;
    e^{\frac{i}{\hbar} \hat{p} x_c(t)}\,
    \hat{p}\,
    e^{-\frac{i}{\hbar} \hat{p} x_c(t)}
    = \hat{p}\,.
    \label{eq:kh_translation_action}
\end{equation}
The second factor in Eq.~\eqref{eq:kh_vector_transformation} is a
spatially uniform phase. Its time derivative cancels the residual
field-dependent scalar term generated by the kinetic energy in
\(\hat{H}_{\mbox{\scriptsize{v}}}(t)\). Indeed,
\begin{equation}
    \hat{U}_{\mbox{\scriptsize{KH}}\leftarrow{\mbox{\scriptsize{v}}}}(t)\,
    \hat{H}_{\mbox{\scriptsize{v}}}(t)\,
    \hat{U}_{\mbox{\scriptsize{KH}}\leftarrow{\mbox{\scriptsize{v}}}}^{\dagger}(t)
    =
    \frac{\left[\hat{p} + p_c(t)\right]^2}{2 m}
    -
    g\,\delta\!\left[\hat{x} + x_c(t)\right]\,,
    \label{eq:kh_transformed_vector_hamiltonian}
\end{equation}
whereas
\begin{equation}
    i \hbar\,
    \dot{\hat{U}}_{\mbox{\scriptsize{KH}}\leftarrow{\mbox{\scriptsize{v}}}}(t)\,
    \hat{U}_{\mbox{\scriptsize{KH}}\leftarrow{\mbox{\scriptsize{v}}}}^{\dagger}(t)
    =
    -\frac{\hat{p}\,p_c(t)}{m}
    -
    \frac{p_c^2(t)}{2 m}\,.
    \label{eq:kh_time_derivative_term}
\end{equation}
Combining Eqs.~\eqref{eq:kh_transformed_vector_hamiltonian} and
\eqref{eq:kh_time_derivative_term} gives
\begin{equation}
    \hat{H}_{\mbox{\scriptsize{KH}}}(t)
    =
    \frac{\hat{p}^2}{2 m}
    -
    g\,\delta\!\left[\hat{x} + x_c(t)\right]\,.
    \label{eq:kh_hamiltonian}
\end{equation}
Thus, the external field is removed from the kinetic term and
absorbed entirely into the time-dependent displacement of the
binding potential, so that the delta-function contact is located at
\(x = -x_c(t)\).

The direct transformation from the scalar-potential gauge to the KH
representation is accordingly
\begin{equation}
    \hat{U}_{\mbox{\scriptsize{KH}}\leftarrow{\mbox{\scriptsize{s}}}}(t)
    =
    \hat{U}_{\mbox{\scriptsize{KH}}\leftarrow{\mbox{\scriptsize{v}}}}(t)\,
    \hat{U}_{\mbox{\scriptsize{v}}\leftarrow\mbox{\scriptsize{s}}}(t)\,,
    \label{eq:scalar_to_kh_operator}
\end{equation}
where
\begin{equation}
    \hat{U}_{\mbox{\scriptsize{v}}\leftarrow\mbox{\scriptsize{s}}}(t)
    =
    \exp\!\left[
        -\frac{i}{\hbar} \hat{x} p_c(t)
    \right].
    \label{eq:scalar_to_vector_operator}
\end{equation}
Hence
\begin{equation}
    |\psi_{\mbox{\scriptsize{KH}}}(t)\rangle
    =
    \hat{U}_{\mbox{\scriptsize{KH}}\leftarrow{\mbox{\scriptsize{s}}}}(t)\,
    |\psi_{\mbox{\scriptsize{s}}}(t)\rangle\,.
    \label{eq:scalar_to_kh_state}
\end{equation}
Because \(p_c(0) = x_c(0) = s_c(0) = 0\), all gauge and KH
transformations reduce to the identity at the initial time. The
laboratory survival amplitude in
Eq.~\eqref{eq:physical_survival_amplitude} may therefore be written
in the KH representation as
\begin{equation}
    a_b(t)
    =
    \langle\psi_b|
    \hat{U}_{\mbox{\scriptsize{s}}\leftarrow\mbox{\scriptsize{KH}}}(t)\,
    \hat{U}_{\mbox{\scriptsize{KH}}}(t,0)
    |\psi_b\rangle
    =
    \langle\psi_b|
    \hat{U}_{\mbox{\scriptsize{s}}\leftarrow\mbox{\scriptsize{KH}}}(t)
    |\psi_{\mbox{\scriptsize{KH}}}(t)\rangle
    =
    \langle\chi_b(t)|\psi_{\mbox{\scriptsize{KH}}}(t)\rangle\,,
    \label{eq:survival_amplitude_kh_operator}
\end{equation}
where
$\hat{U}_{\mbox{\scriptsize{s}}\leftarrow\mbox{\scriptsize{KH}}}(t)
=
\hat{U}_{\mbox{\scriptsize{KH}}\leftarrow\mbox{\scriptsize{s}}}^{\dagger}(t)$,
and $|\chi_b(t)\rangle =
\hat{U}_{\mbox{\scriptsize{KH}}\leftarrow\mbox{\scriptsize{s}}}(t)
|\psi_b\rangle$ is the field-free bound state represented in the KH
picture at the observation time. In the coordinate representation,
this state is given by
\begin{equation}\label{eq:bound_state_in_KH}
    \chi_b(x,t)
    =
    e^{\frac{i}{\hbar} s_c(t)}\,
    e^{-\frac{i}{\hbar} p_c(t)\,\left[x + x_c(t)\right]}\,
    \psi_b\!\left(x + x_c(t)\right)\,.
\end{equation}
Eq.~\eqref{eq:survival_amplitude_kh_operator} shows explicitly that
the physical survival amplitude is not simply the instantaneous
overlap \(\langle\psi_b|\psi_{\mbox{\scriptsize{KH}}}(t)\rangle\).
Rather, the final field-free bound-state bra must first be
transformed into the KH representation in order to represent the
same laboratory measurement.

\subsection{Exact Moving-Contact Integral Equation}
\label{subsec:moving_contact}
In the KH representation, the delta-function potential is displaced
to the moving position \(-x_c(t)\). For the KH Hamiltonian in
Eq.~\eqref{eq:kh_hamiltonian}, the exact Lippmann--Schwinger
equation for the wavefunction takes the form
\begin{equation}
    \psi_{\mbox{\scriptsize{KH}}}(x,t)
    =
    \phi_{\mbox{\scriptsize{KH}}}(x,t)
    +
    \frac{i g}{\hbar}
    \int_0^t d\tau\,
    K_0\!\left(
        x,t\middle|-x_c(\tau),\tau
    \right)\,
    \psi_{\mbox{\scriptsize{KH}}}
    \!\left(
        -x_c(\tau),\tau
    \right)\,,
    \label{eq:kh_lippmann_schwinger_wavefunction}
\end{equation}
where
\begin{equation}
    \phi_{\mbox{\scriptsize{KH}}}(x,t)
    =
    \int_{-\infty}^{\infty}
    dx'\,
    K_0(x,t|x',0)\,
    \psi_b(x')
    \label{eq:kh_homogeneous_wavefunction}
\end{equation}
represents the homogeneous KH solution generated by propagating the
initial bound state with \(K_0\). Here, \(K_0\) denotes the
one-dimensional free-particle propagator. In the KH representation,
it describes propagation between contact encounters, while the
influence of the external field is retained through the trajectory
of the moving contact. Its explicit form evaluated with the source
point at the moving contact, \(x' = -x_c(\tau)\), is
\begin{equation}
    K_0\!\left(
        x,t\middle|-x_c(\tau),\tau
    \right)
    =
    \sqrt{
        \frac{m}
        {2\pi i\hbar (t - \tau)}
    }\,
    \exp\!\left[
        \frac{
            im\left[
                x + x_c(\tau)
            \right]^2
        }
        {
            2 \hbar (t - \tau)
        }
    \right]\,.
    \label{eq:kh_free_propagator}
\end{equation}

Evaluating Eq.~\eqref{eq:kh_lippmann_schwinger_wavefunction} at the
moving contact position yields the closed Volterra integral equation
for the corresponding contact amplitude, \(q(t)\equiv
\psi_{\mbox{\scriptsize{KH}}}\!\left(-x_c(t),t\right)\), given by
\begin{equation}
    q(t)
    =
    q_0(t)
    +
    \frac{i g}{\hbar}
    \int_0^t d\tau\,
    K_c(t,\tau)\,
    q(\tau)\,,
    \label{eq:kh_contact_equation}
\end{equation}
where \(q_0(t) = \phi_{\mbox{\scriptsize
KH}}\!\left(-x_c(t),t\right)\) denotes the homogeneous contribution,
and \(K_c(t,\tau) =
K_0\!\left(-x_c(t),t\middle|-x_c(\tau),\tau\right)\) is the
corresponding moving-contact kernel, obtained by substituting \(x =
-x_c(t)\) into Eq.~\eqref{eq:kh_free_propagator}. For the dc
trajectory, one finds
\begin{equation}
    x_c(t) - x_c(\tau)
    =
    \frac{F_0}{2 m}
    \left(t^2 - \tau^2\right)
    =
    \frac{F_0}{2 m}\,
    (t - \tau)(t + \tau)\,,
    \label{eq:dc_displacement_difference}
\end{equation}
so that the contact kernel becomes
\begin{equation}
    K_c(t,\tau)
    =
    \sqrt{
        \frac{m}
        {2\pi i \hbar (t - \tau)}
    }\,
    \exp\!\left[
        \frac{i F_0^2}
        {8 m \hbar}\,
        (t - \tau)(t + \tau)^2
    \right]\,.
    \label{eq:dc_kh_contact_kernel}
\end{equation}
For later use, the homogeneous contact amplitude appearing in
Eq.~\eqref{eq:kh_contact_equation} may be written explicitly as
\begin{equation}\label{eq:q_0-1}
    q_0(t)
    =
    \sqrt{\frac{\kappa\,m}{2 \pi i \hbar t}}
    \sum_{\sigma = \pm 1}
    I_\sigma\!\left(t;-x_c(t),0\right)\,,
\end{equation}
where the generalized form is
\begin{equation}\label{eq:I_sigma_q0}
    I_\sigma(t;\Delta,p)
    =
    \frac{\sqrt{\pi}}
    {2 \sqrt{-i \beta_t}}\,
    \exp\!\left[
        i \beta_t \Delta^2
        -
        \frac{B_\sigma^2(t;\Delta,p)}
        {4 i \beta_t}
    \right]
    \erfc\!\left[
        -\frac{B_\sigma(t;\Delta,p)}
        {2 \sqrt{-i \beta_t}}
    \right]
\end{equation}
with
\begin{equation}\label{eq:B_sigma_general}
    \beta_t = \frac{m}{2 \hbar t}\;\;\; ;\;\;\;
    B_\sigma(t;\Delta,p)
    =
    -\kappa
    +
    \frac{i \sigma p}{\hbar}
    -
    2 i \sigma \beta_t \Delta\,.
\end{equation}
The derivation is given in Appendix~\ref{app:q0_derivation}.

Once Eq.~\eqref{eq:kh_contact_equation} has been solved for
\(q(t)\), substitution into
Eq.~\eqref{eq:kh_lippmann_schwinger_wavefunction} gives the exact KH
wavefunction \(\psi_{\mbox{\scriptsize KH}}(x,t)\). A further
substitution into Eq.~\eqref{eq:survival_amplitude_kh_operator}
yields the exact decomposition
\begin{equation}
    a_b(t)
    =
    a_{b}^{(0)}(t)
    + a_{b}^{(\delta)}(t)\,,
    \label{eq:kh_survival_amplitude_contact_form}
\end{equation}
where
\begin{equation}
    a_{b}^{(0)}(t)
    =
    \int dx\, \chi_b^{\ast}(x,t)\, \phi_{\mbox{\scriptsize
    KH}}(x,t)\;\;\; ;\;\;\;
    a_{b}^{(\delta)}(t)
    =
    \frac{i g}{\hbar}
    \int_0^t d\tau\,
    \mathcal{J}_b(t,\tau)\,
    q(\tau)\,,
    \label{eq:kh_direct_amplitude}
\end{equation}
with the bound-state contact kernel
\begin{equation}
    \mathcal{J}_b(t,\tau)
    =
    \int_{-\infty}^{\infty}dx\,
    \chi_b^{\ast}(x,t)\,
    K_0\!\left(
        x,t
        \middle|
        -x_c(\tau),\tau
    \right)\,.
    \label{eq:kh_bound_contact_kernel}
\end{equation}
Eqs.~\eqref{eq:kh_contact_equation} and
\eqref{eq:kh_survival_amplitude_contact_form} therefore provide an
exact contact-level representation of the physical bound-state
dynamics. Within the KH dynamical equation, the external field
enters through the classical trajectory \(x_c(t)\), while all
repeated interactions with the delta-function well are encoded in
the contact amplitude \(q(t)\). No approximation has been introduced
in obtaining this reduction.

\subsection{Remaining Analytical Challenge}
\label{subsec:remaining_structure}
The exact KH formulation in terms of the contact amplitude \(q(t)\)
reduces the driven dynamics to free-particle propagation between
successive interactions with the moving contact. Nevertheless, the
contact dynamics itself does not yet possess a simple convolution
structure in laboratory time. For the uniform dc field, the
moving-contact kernel in Eq.~\eqref{eq:dc_kh_contact_kernel}
contains a highly oscillatory phase that depends on both the
relative-time combination \((t - \tau)\) and the complementary
combination \((t + \tau)\). By contrast, the scalar-gauge contact
kernel discussed in Subsec.~\ref{subsec:vector_gauge} contains the
cubic phase in Eq.~\eqref{eq:scalar_cubic_phase}, which depends only
on the relative time \((t - \tau)\). Consequently, the KH
moving-contact kernel \(K_c(t,\tau)\) is not initially a function of
the chronological time difference alone, and the physical Volterra
equation \eqref{eq:kh_contact_equation} cannot be reduced directly
to an algebraic equation through the ordinary Laplace convolution
theorem.

The exact contact formulation nevertheless provides the essential
reduction: the remaining analytical difficulty no longer lies in
constructing the complete chronological propagator or wavefunction,
but in reorganizing the two-time structure of the contact-level
dynamics identified above. The relative time \((t - \tau)\)
specifies the duration of a propagation interval, while \((t +
\tau)/2\) specifies its mean temporal location. A reorganization
that treats these complementary temporal roles separately is
therefore more suitable for extracting the physical bound-state
amplitude. To implement this reorganization at the level of the
bound-state projection, Sec.~\ref{sec:uw} begins by introducing the
accumulated bound-state amplitude
\begin{equation}
    A_b(t)
    =
    \int_0^t ds\,a_b(s)\,,
    \label{eq:accumulated_amplitude_preview}
\end{equation}
from which the physical amplitude is recovered through
\begin{equation}
    a_b(t)
    =
    \frac{d A_b(t)}{dt}\,.
    \label{eq:physical_from_accumulated_preview}
\end{equation}
The additional time integration over \(s\) enlarges the temporal
description and permits an exact transformation from the original
chronological pair \((s,\tau)\) to the relative- and center-time
variables. Here, the relative time specifies the duration of a
propagation interval, while the center time specifies its mean
temporal location. This reorganization makes the corresponding
interval geometry explicit. As shown in
Sec.~\ref{sec:exact_solution}, when the chronological expansion of
the physical Volterra equation is inserted into the bound-state
projection, the dependence on the intermediate center times cancels
exactly, leaving an ordinary convolution hierarchy in the successive
relative-time intervals.

\section{Temporal Reorganization in Relative- and Center-Time Variables}
\label{sec:uw}
%
\subsection{Relative- and Center-Time Variables}
The accumulated bound-state formulation introduced at the end of
Sec.~\ref{sec:kh_formulation} naturally gives rise to an exact
two-time description involving the interaction time \(\tau\) and the
observation time \(s\). To reorganize this temporal structure, we
introduce the relative- and center-time variables
\begin{subequations}
\begin{equation}
    u = s - \tau\;\;\; ;\;\;\;
    w = \frac{s + \tau}{2}\,,
    \label{eq:uw_definition}
\end{equation}
or equivalently,
\begin{equation}
    \tau = w - \frac{u}{2}\;\;\; ;\;\;\;
    s = w + \frac{u}{2}\,.
    \label{eq:uw_inverse}
\end{equation}
\end{subequations}
This transformation is exact and has unit Jacobian,
\begin{equation}
    \left|
    \frac{\partial(\tau,s)}
         {\partial(u,w)}
    \right|
    = 1\,,
    \label{eq:uw_jacobian}
\end{equation}
so that $ds\,d\tau = du\,dw$.

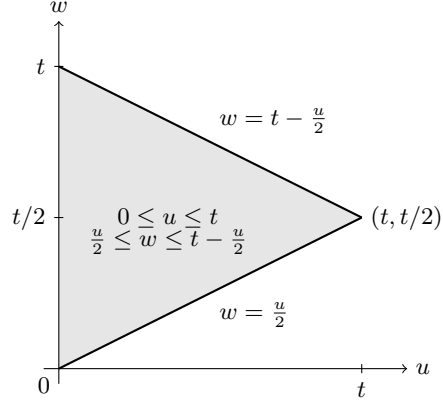
\begin{figure}[t] \centering
\begin{tikzpicture}[scale=4.0]
%
\def\t{1}
%
\fill[gray!20] (0,0) -- (1,0.5) -- (0,1) -- cycle;
%
\draw[thick] (0,0) -- (1,0.5) node[midway, below right]
{$w=\frac{u}{2}$}; \draw[thick] (0,1) -- (1,0.5) node[midway, above
right] {$w=t-\frac{u}{2}$};
%
\draw[->] (-0.05,0) -- (1.15,0) node[right] {$u$}; \draw[->]
(0,-0.05) -- (0,1.15) node[above] {$w$};
%
\draw (1,0.015) -- (1,-0.015) node[below] {$t$}; \draw (0.015,1) --
(-0.015,1) node[left] {$t$}; \draw (0.015,0.5) -- (-0.015,0.5)
node[left] {$t/2$};
%
\node[below left] at (0,0) {$0$}; \node[right] at (1,0.5)
{$(t,t/2)$};
%
\node at (0.36,0.50) {$0\le u\le t$}; \node at (0.36,0.42)
{$\frac{u}{2}\le w\le t-\frac{u}{2}$};
\end{tikzpicture}
\caption{Integration region in the \((u,w)\) plane. The original
chronological domain \((0 \leq \tau \leq s \leq t)\) is mapped to
\((0 \leq u \leq t)\) and \((u/2 \leq w \leq t-u/2)\). When the
order of integration is reversed, the fixed-\(w\) horizontal slices
have the upper limit \(u = 2 \min(w,t-w)\), corresponding to \(u = 2
w\) for \(0 \leq w \leq t/2\) and \(u = 2 (t-w)\) for \(t/2 \leq w
\leq t\).} \label{fig:uw-integration-region}
\end{figure}

The original triangular integration domain, defined by \(0 \le \tau
\le s \le t\), is thereby transformed into
\begin{subequations}
\begin{equation}
    0 \le u \le t\;\;\; ;\;\;\;
    \frac{u}{2}
    \le
    w
    \le
    t - \frac{u}{2}\,,
    \label{eq:uw_domain}
\end{equation}
or equivalently,
\begin{equation}
    \int_0^t ds
    \int_0^s d\tau
    =
    \int_0^t du
    \int_{u/2}^{\,t - u/2} dw\,.
    \label{eq:uw_domain_integral}
\end{equation}
\end{subequations}
This transformation of the integration domain is illustrated in
Fig.~\ref{fig:uw-integration-region}. The order of integration over
\(u\) and \(w\) can also be reversed. From Eq.~\eqref{eq:uw_domain},
solving the two inequalities for \(u\) gives \(u \le 2 w\) and \(u
\le 2 (t-w)\). Together with \(0 \le u\), these conditions yield
\begin{subequations}
\begin{equation}\label{eq:wu_domain}
    0 \le u \le 2\,\min(w,t-w)\,.
\end{equation}
Accordingly, Eq.~\eqref{eq:uw_domain_integral} can be rewritten as
\begin{equation}\label{eq:wu_domain_integral0}
    \int_0^t\,du \int_{u/2}^{t-u/2}\,dw
    = \int_0^t\,dw \int_0^{2\,\min(w,t-w)}\,du =
    \int_0^{t/2}\,dw \int_0^{2w}\,du
    +
    \int_{t/2}^{t}\,dw \int_0^{2(t-w)}\,du\,.
\end{equation}
\end{subequations}
The transformation to \((u,w)\) is not merely a change of
integration variables. It reorganizes the original chronological
two-time description in terms of variables having complementary
temporal roles. The relative time \(u\) measures the temporal
separation between the interaction and observation events, whereas
the center time \(w\) specifies their mean temporal location. This
exact temporal reorganization provides a geometrical framework for
organizing the propagation intervals entering the accumulated
bound-state amplitude. Its implications for the complete
chronological rescattering dynamics are developed in
Sec.~\ref{sec:exact_solution}.

\subsection{Structural Consequences of the Temporal Reorganization}
\label{subsec:3-2}
The structural significance of this temporal reorganization becomes
most apparent when the transformed domain is interpreted as a space
of propagation intervals rather than merely as a new integration
region. A single interval refers to propagation from an earlier
contact-interaction time \(\tau\) to a later observation time \(s\).
According to Eq.~\eqref{eq:uw_definition}, the pair \((\tau,s)\)
uniquely determines \((u,w)\); hence, each such interval is
represented by a unique point \((u,w)\) in the triangular domain of
Fig.~\ref{fig:uw-integration-region}. Conversely,
Eq.~\eqref{eq:uw_inverse} uniquely recovers the two endpoints
\(\tau\) and \(s\) from that point. When the center time is fixed at
\(w = w_0\), varying \(u\) traces a horizontal segment across this
domain. Each point on that segment represents a different
propagation interval having the same temporal midpoint \(w_0\), with
its two endpoints determined by Eq.~\eqref{eq:uw_inverse} evaluated
at \(w = w_0\). Along this horizontal segment, the allowed range of
\(u\) is given by Eq.~\eqref{eq:wu_domain}. For example, at \(w_0 =
t/2\), the interval expands symmetrically from the degenerate
interval \(\tau = s = t/2\) at \(u = 0\) to the full interval
\([0,t]\) at \(u = t\).

The fixed-\(w\) horizontal slices shown in
Fig.~\ref{fig:uw-integration-region} provide the geometrical basis
for interchanging the order of integration, as expressed in
Eq.~\eqref{eq:wu_domain_integral0}. In particular, they decompose
the triangular domain into the two sectors \(0 \leq w \leq t/2\) and
\(t/2 \leq w \leq t\), with the corresponding upper limits \(2w\)
and \(2 (t-w)\) for the relative time \(u\). In the original
\(u\)-first representation of Eq.~\eqref{eq:uw_domain_integral}, the
\(t\)-dependent outer boundary is \(w = t - u/2\). After the order
of integration is interchanged, the two sloping boundaries of the
triangular domain appear as the upper \(u\)-boundaries of the two
sectors. In the first sector \(0 \leq w \leq t/2\), the boundary \(u
= 2w\) corresponds, through Eq.~\eqref{eq:uw_inverse}, to \((\tau =
0; s = 2w)\), and therefore represents the initial interaction-time
boundary of the original chronological domain. In the second sector
\(t/2 \leq w \leq t\), the boundary \(u = 2 (t-w)\) corresponds to
\((\tau = 2w - t; s = t)\), and therefore represents the final
observation-time boundary. The two sectors meet along the internal
dividing line \(w = t/2\).

As shown in Sec.~\ref{sec:exact_solution}, this reorganization plays
an essential role in evaluating the accumulated contact contribution
\(A_b^{(\delta)}(t)\). Upon differentiation of \(A_b^{(\delta)}(t)\)
with respect to \(t\), the boundary terms generated by the above two
sectors at the moving internal dividing line \(w=t/2\) cancel
exactly. The boundary \(u = 2w\), corresponding to \(\tau = 0\), has
no explicit \(t\)-dependence and produces no remaining boundary
contribution. The surviving term is therefore a one-dimensional
integral along the boundary \(u = 2 (t-w)\), corresponding to \(s =
t\). Reparametrizing this boundary by \(u\), so that \(w = t -
u/2\), gives \(\tau = t - u\) and \(s = t\). The resulting
one-dimensional integral for the physical contact contribution
\(a_b^{(\delta)}(t)\) consequently contains the physical contact
amplitude \(q(\tau)\) evaluated at \(\tau = t - u\), namely
\(q(t-u)\).

For later use, we emphasize that the convolution structure of the
contact dynamics has a distinct origin and should not be attributed
to integration along a fixed center-time slice. As shown in
Sec.~\ref{sec:exact_solution}, the moving-contact kernel
\(K_c(t,\tau)\) acting on \(q(\tau)\) in
Eq.~\eqref{eq:kh_contact_equation} propagates the contact amplitude
from the earlier contact time \(\tau\) to the later contact time
\(t\). For each such propagation segment, with its own center time,
the kernel can be factorized directly in the original chronological
variables into two endpoint phase factors and a residual
relative-time kernel factor depending only on the duration of that
segment. When the kernels associated with successive propagation
segments are multiplied to form a complete chronological history,
the adjacent endpoint phase factors cancel pairwise at all
intermediate contact times. As a result, the intermediate
center-time dependence disappears from the complete chronological
history. Remarkably, the remaining relative-time kernel factors
depend only on the durations of the corresponding propagation
intervals, allowing the complete rescattering hierarchy to be
organized as an ordinary convolution over those durations.

\section{Exact Analytical Solution}
\label{sec:exact_solution}
%
We now exploit the temporal reorganization developed in
Sec.~\ref{sec:uw} to derive the exact finite-time bound-state
amplitude. The accumulated amplitude serves as the analytical
intermediary through which the chronological contact dynamics can be
reorganized. We begin by decomposing it into a contact-free
contribution and a delta-interaction contribution.

\subsection{Accumulated Bound-State Amplitude in Relative- and Center-Time}
The accumulated bound-state amplitude introduced in
Eq.~\eqref{eq:accumulated_amplitude_preview} can be written as
\begin{equation}
    A_b(t)
    =
    A_b^{(0)}(t)
    +
    A_b^{(\delta)}(t)\,,
    \label{eq:Ab_decomposition}
\end{equation}
where the contact-free contribution is
\begin{equation}
    A_b^{(0)}(t)
    =
    \int_0^t ds\,a_b^{(0)}(s)\,,
    \label{eq:Ab_free_definition}
\end{equation}
with \(a_b^{(0)}(s)\) obtained directly from
Eqs.~\eqref{eq:bound_state_in_KH},
\eqref{eq:kh_homogeneous_wavefunction}, and
\eqref{eq:kh_direct_amplitude}. This contact-free contribution can
be evaluated explicitly and will be reduced to closed form below.
Likewise, the nontrivial contact contribution is
\begin{equation}
    A_b^{(\delta)}(t)
    =
    \int_0^t ds\,a_b^{(\delta)}(s)\,,
    \label{eq:Ab_delta_definition}
\end{equation}
where \(a_b^{(\delta)}(s)\) follows directly from
Eqs.~\eqref{eq:kh_direct_amplitude} and
\eqref{eq:kh_bound_contact_kernel}. Unlike the contact-free
contribution \(A_b^{(0)}(t)\), the exact evaluation of
\(A_b^{(\delta)}(t)\) must retain the complete history of repeated
interactions with the moving contact. In the KH representation, the
field-dependent phase associated with an individual propagation
interval takes a particularly simple form: when expressed in terms
of its relative time and center time, it is linear in the relative
time at fixed center time. This single-interval simplification does
not by itself make the physical contact equation convolutional. Its
significance emerges when the complete chronological rescattering
expansion is assembled, as developed below.

Using the relative- and center-time variables \((u,w)\) introduced
in Sec.~\ref{sec:uw}, we apply the transformed integration domain
given in Eq.~\eqref{eq:uw_domain_integral} and interchange the order
of integration according to Eq.~\eqref{eq:wu_domain_integral0}. The
contact contribution in Eq.~\eqref{eq:Ab_delta_definition} can then
be written explicitly as
\begin{subequations}
\begin{equation}\label{eq:A_b_delta0}
    A_b^{(\delta)}(t)
    =
    \frac{i g}{\hbar}
    \int dx
    \left[
    \int_0^{t/2} dw \int_0^{2w} du
    +
    \int_{t/2}^{t} dw \int_0^{2(t-w)} du
    \right]\,
    \mathcal{F}_b(x,u,w)\;
    q\!\left(w - \frac{u}{2}\right)\,,
\end{equation}
where
\begin{equation}
\begin{aligned}
    \mathcal{F}_b(x,u,w)
    ={}&
    \exp\!\left[
        -\frac{i}{\hbar}
        \left\{
            s_c\!\left(w + \frac{u}{2}\right)
            -
            p_c\!\left(w + \frac{u}{2}\right)
            \left[
                x + x_c\!\left(w + \frac{u}{2}\right)
            \right]
        \right\}
    \right]\;
    \psi_b^*\!\left[
        x + x_c\!\left(w + \frac{u}{2}\right)
    \right]
    \times
    \\
    &\sqrt{\frac{m}{2 \pi i\hbar u}}\,
    \exp\!\left[
        \frac{im}{2 \hbar u}
        \left[
            x + x_c\!\left(w - \frac{u}{2}\right)
        \right]^2
    \right]\,.
\end{aligned}
\label{eq:F_uw}
\end{equation}
\end{subequations}
The detailed derivation of Eq.~\eqref{eq:A_b_delta0} is presented in
Appendix~\ref{app:uw_derivation}. Equation~\eqref{eq:A_b_delta0} is
an exact rewriting of the accumulated contact contribution. For each
fixed value of \(w\) within the inner integration, \(u\)
parametrizes the corresponding relative-time interval. This is a
kinematic organization of the two-time integration domain. It is
important, however, to distinguish this fixed-\(w\) integration from
the chronological dynamics governed by the physical Volterra
equation \eqref{eq:kh_contact_equation}. In that equation, for a
fixed observation time \(t\), varying the interaction time \(\tau\)
changes both variables \((u,w)\) in Eq.~\eqref{eq:uw_definition}
simultaneously. Thus, the center time is \emph{not} fixed along the
chronological integration defining the physical contact amplitude;
indeed, for a fixed observation time \(t\), it varies according to
\(w = (t + \tau)/2\) as \(\tau\) runs from \(0\) to \(t\).
Accordingly, \(q(w - u/2)\) in Eq.~\eqref{eq:A_b_delta0} remains the
physical contact amplitude \(q(\tau)\), with \(\tau = w - u/2\),
determined by Eq.~\eqref{eq:kh_contact_equation} and containing its
complete chronological rescattering history.

For subsequent use, the \(x\)-integral of \(\mathcal{F}_b\) in
Eq.~\eqref{eq:F_uw} can also be evaluated explicitly. Following the
same half-line decomposition used for \(q_0(t)\) in
Eq.~\eqref{eq:q_0-1}, whose detailed derivation is given in
Appendix~\ref{app:q0_derivation}, and using the generalized form in
Eq.~\eqref{eq:I_sigma_q0}, we obtain
\begin{equation}\label{eq:X-final}
\begin{aligned}
    X_b(u,w)
    &\equiv
    \int_{-\infty}^{\infty} dx\,
    \mathcal{F}_b(x,u,w)
    =
    \sqrt{\frac{\kappa\,m}{2 \pi i \hbar u}}\,
    \exp\left[-\frac{i}{\hbar} s_c(w + u/2)\right]
    \sum_{\sigma = \pm 1}
    I_\sigma\!\left(
        u;\Delta(u,w),p_c(w + u/2)
    \right)\,,
\end{aligned}
\end{equation}
where
\begin{equation}\label{eq:Delta_uw}
    \Delta(u,w)
    =
    x_c\!\left(w + \frac{u}{2}\right)
    -
    x_c\!\left(w - \frac{u}{2}\right)
    =
    \frac{F_0\,w\,u}{m}\,.
\end{equation}

\subsection{Chronological Rescattering and the Exact Relative-Time
Convolution Hierarchy}
As emphasized in Subsec.~\ref{subsec:3-2}, the fixed-\(w\)
description of a single propagation interval in
Eq.~\eqref{eq:A_b_delta0} is a kinematic organization of the
accumulated-amplitude integration domain and must not be interpreted
as a fixed-\(w\) evolution equation for the physical contact
amplitude \(q(t)\). In Eq.~\eqref{eq:A_b_delta0}, this amplitude is
evaluated at the interaction time \(\tau = w - u/2\) and therefore
appears simply as \(q(\tau) = q(w - u/2)\). By contrast, its
dynamics remains governed by the chronological Volterra equation
\eqref{eq:kh_contact_equation}, in which successive propagation
intervals generally have different center times. The convolution
structure of the complete contact dynamics emerges from the exact
factorization of the moving-contact kernels appearing in the
chronological rescattering expansion of that equation.

To expose this structure, we note that the elementary identity
\begin{equation}
    (t - \tau) (t + \tau)^2
    =
    \frac{4}{3} \left(t^3 - \tau^3\right)
    -
    \frac{1}{3} (t - \tau)^3
    \label{eq:dc_phase_identity}
\end{equation}
permits the moving-contact kernel \eqref{eq:dc_kh_contact_kernel} to
be factorized exactly as
\begin{equation}
    K_c(t,\tau)
    =
    e^{i \Theta(t)}\,
    k_{\scriptstyle{F}}(t - \tau)\,
    e^{-i \Theta(\tau)}\,,
    \label{eq:Kc_factorization}
\end{equation}
where the single-interval relative-time kernel is
\begin{equation}
    k_{\scriptstyle{F}}(u)
    =
    \sqrt{\frac{m}{2 \pi i \hbar u}}\,
    \exp\!\left(-i \gamma u^3\right)\,,
    \qquad
    u > 0
    \label{eq:kF_definition}
\end{equation}
with
\[
    \gamma
    =
    \frac{F_0^2}{24 m \hbar}\;\;\; ;\;\;\;
    \Theta(t)
    =
    4 \gamma t^3
    =
    \frac{F_0^2 t^3}{6 m \hbar}\,.
\]
Thus, the center-time dependence of an individual KH propagation
interval has not been discarded; rather, it has been separated into
the two phase factors \(e^{i \Theta(t)}\) and \(e^{-i
\Theta(\tau)}\), associated respectively with the later and earlier
endpoints of that interval, while the residual kernel
\(k_{\scriptstyle{F}}(t - \tau)\) depends only on its duration \(u =
t - \tau\).

This separation becomes decisive for a complete chronological
history. For \(n \geq 1\), let \(t = t_0 > t_1 > \cdots > t_n \geq
0\), and introduce the successive relative-time intervals
\begin{equation}
    u_j = t_j - t_{j+1} > 0\,,
    \qquad
    j = 0, 1, \cdots, n-1\,.
    \label{eq:successive_intervals}
\end{equation}
Repeated use of Eq.~\eqref{eq:Kc_factorization} gives
\begin{equation}
    \prod_{j=0}^{n-1} K_c(t_j,t_{j+1})
    =
    e^{i \Theta(t_0)}
    \left[
        \prod_{j=0}^{n-1} k_{\scriptstyle{F}}(u_j)
    \right]
    e^{-i \Theta(t_n)}\,.
\label{eq:chronological_phase_cancellation}
\end{equation}
The endpoint phase factors associated with the intermediate contact
times \(t_1, \cdots, t_{n-1}\) cancel pairwise. Consequently, the
center-time dependence of the successive propagation intervals
disappears from the complete product, leaving only the two overall
endpoint phase factors and the relative-time kernel factors
\(k_{\scriptstyle{F}}(u_j)\), each of which depends only on the
duration \(u_j\) of the corresponding interval.

At the same time, the change from the ordered interaction times
\(t_1, \cdots, t_n\) to the successive positive interval durations
\(u_0, \cdots, u_{n-1}\) transforms the chronologically ordered
integration domain \(0\leq t_n\leq\cdots\leq t_1\leq t\) into the
standard simplex
\[
    u_j \geq 0\,,
    \qquad
    \sum_{j=0}^{n-1} u_j \leq t\,,
\]
which is precisely the integration domain underlying an iterated
time convolution. Eq.~\eqref{eq:chronological_phase_cancellation}
therefore identifies the exact origin of the ordinary relative-time
convolution hierarchy: it arises from the pairwise cancellation of
the intermediate endpoint phase factors together with the resulting
dependence on the successive interval durations alone, rather than
from holding one common center time fixed throughout the physical
evolution.

Equivalently, the endpoint-phase cancellation can be incorporated
directly into the physical contact equation by factoring the overall
endpoint phase out of the contact amplitudes: \(Q(t) = e^{-i
\Theta(t)} q(t)\) and \(Q_0(t) = e^{-i \Theta(t)} q_0(t)\).
Multiplying Eq.~\eqref{eq:kh_contact_equation} by \(e^{-i
\Theta(t)}\) and using Eq.~\eqref{eq:Kc_factorization} then yields
the exact convolution equation
\begin{equation}
    Q(t)
    =
    Q_0(t)
    +
    \frac{i g}{\hbar}
    \int_0^t d\tau\,
    k_{\scriptstyle{F}}(t-\tau)\,Q(\tau)\,.
    \label{eq:Q_convolution}
\end{equation}
This rephased equation is simply the compact form of the telescoping
cancellation in Eq.~\eqref{eq:chronological_phase_cancellation}; no
change has been made to the physical Volterra dynamics.

For functions supported on the nonnegative time axis, let \(*\)
denote ordinary Volterra convolution,
\begin{equation}
    (f*g)(t)
    =
    \int_0^t d\tau\,
    f(t-\tau)\,g(\tau)\,.
    \label{eq:Volterra_convolution_definition}
\end{equation}
The notation \(k_{\scriptstyle{F}}^{*n}\) denotes the \(n\)-fold
convolution power of \(k_{\scriptstyle{F}}\), rather than complex
conjugation. In particular,
\begin{equation}
    k_{\scriptstyle{F}}^{*0}(t) = \delta(t)\;\;\; ;\;\;\;
    k_{\scriptstyle{F}}^{*1}(t) = k_{\scriptstyle{F}}(t)\;\;\; ;\;\;\;
    k_{\scriptstyle{F}}^{*2}(t)
    =
    (k_{\scriptstyle{F}}*k_{\scriptstyle{F}})(t)
    =
    \int_0^t d\tau\,
    k_{\scriptstyle{F}}(t-\tau)\,k_{\scriptstyle{F}}(\tau)\,.
\label{eq:kF_convolution_powers}
\end{equation}
Here \(k_{\scriptstyle{F}}^{*0}(t) = \delta(t)\) is the convolution
identity, so that \(\left(k_{\scriptstyle{F}}^{*0}*Q_0\right)(t)
    =
    \left(\delta*Q_0\right)(t)
    =
    Q_0(t)\).
Iteration of Eq.~\eqref{eq:Q_convolution} therefore gives the
all-orders chronological rescattering expansion
\begin{equation}
    Q(t)
    =
    \sum_{n=0}^{\infty}
    \left(\frac{i g}{\hbar}\right)^n
    \left(k_{\scriptstyle{F}}^{*n}*Q_0\right)(t)\,.
    \label{eq:Q_rescattering_hierarchy}
\end{equation}
Here \(n\) counts the number of internal contact-to-contact
rescattering events. In particular, the \(n=0\) term is simply
\(Q_0(t)\), the zeroth-order contact amplitude, which contains no
internal contact-to-contact rescattering and generates the direct
contact contribution when inserted into the final bound-state
projection. This counting convention is the one used in
Fig.~\ref{fig:bound_state_probability}.

The hierarchy may equivalently be collected into the exact
convolution resolvent
\begin{equation}
    \mathcal{R}_{\scriptstyle{F}}(t)
    =
    \delta(t) + R_{\scriptstyle{F}}(t)
    =
    \sum_{n=0}^{\infty}
    \left(\frac{i g}{\hbar}\right)^n
    k_{\scriptstyle{F}}^{*n}(t)\,,
    \label{eq:RF_time_definition}
\end{equation}
so that
\begin{equation}
    q(t)
    =
    e^{i \Theta(t)}
    \left(\mathcal{R}_{\scriptstyle{F}}*Q_0\right)(t)
    =
    q_0(t)
    +
    e^{i \Theta(t)}
    \int_0^t dv\,
    R_{\scriptstyle{F}}(v)\,Q_0(t-v)\,.
    \label{eq:q_resolvent}
\end{equation}
The corresponding Laplace-space representation follows directly from
the convolution theorem, with
\(\mathcal{L}[k_{\scriptstyle{F}}^{*n}](z) =
[\widetilde{k}_{\scriptstyle{F}}(z)]^n\). Indeed, taking the Laplace
transform of Eq.~\eqref{eq:Q_convolution} gives
\[
    \widetilde{Q}(z)
    =
    \widetilde{Q}_0(z)
    +
    \frac{i g}{\hbar}\,
    \widetilde{k}_{\scriptstyle{F}}(z)\,\widetilde{Q}(z)\,,
\]
and hence
\[
    \widetilde{Q}(z)
    =
    \frac{\widetilde{Q}_0(z)}
    {1 - \dfrac{i g}{\hbar}\,\widetilde{k}_{\scriptstyle{F}}(z)}\,.
\]
Equivalently, applying the convolution theorem term by term to
Eq.~\eqref{eq:RF_time_definition} converts its convolution series
into a geometric series. It follows that
\begin{equation}
    \widetilde{R}_{\scriptstyle{F}}(z)
    =
    \frac{
        \dfrac{i g}{\hbar}\,\widetilde{k}_{\scriptstyle{F}}(z)
    }{
        1 - \dfrac{i g}{\hbar}\,\widetilde{k}_{\scriptstyle{F}}(z)
    }\;\;\; ;\;\;\;
    \widetilde{\mathcal{R}}_{\scriptstyle{F}}(z)
    =
    \frac{1}
    {1 - \dfrac{ig}{\hbar}\,\widetilde{k}_{\scriptstyle{F}}(z)}\,.
    \label{eq:RF_laplace}
\end{equation}
From Eq.~\eqref{eq:kF_definition}, the Laplace transform of the
relative-time kernel is
\begin{equation}
    \widetilde{k}_{\scriptstyle{F}}(z)
    =
    \sqrt{\frac{m}{2 \pi i\hbar}}
    \int_0^\infty du\,
    u^{-1/2}
    \exp\!\left(-z u - i \gamma u^3\right)\,,
    \qquad
    \operatorname{Re}z > 0\,.
    \label{eq:kF_laplace}
\end{equation}

The cubic phase in Eq.~\eqref{eq:kF_laplace} is precisely the same
relative-time factor encountered in the conventional scalar-gauge
contact kernel in Eq.~\eqref{eq:scalar_cubic_phase}. The Laplace
transform containing this cubic factor is non-elementary, and the
present temporal reorganization does not remove this intrinsic
dynamical feature of the dc-driven problem. Nor should the
construction be understood as being available only in the KH
representation: by gauge equivalence, the physical bound-state
amplitude and its accumulated form may also be formulated directly
in the scalar gauge. The advantage of the present formulation is
instead structural. At the level of the contact dynamics,
endpoint-phase factorization and rephasing reduce the
nonconvolutional moving-contact kernel to the duration-dependent
kernel \(k_{\scriptstyle F}(t - \tau)\), thereby converting the
physical contact equation into an ordinary Volterra convolution
equation to which the Laplace convolution theorem applies. At the
level of the final bound-state projection, the transformation to
\((u,w)\) provides a transparent geometrical organization of the
accumulated-amplitude integration domain and its subsequent
reduction to the boundary integral that yields the physical contact
contribution \(a_b^{(\delta)}(t)\) to the bound-state amplitude.
These two complementary analytical steps separate the chronological
contact dynamics from the final projection without altering their
gauge-equivalent physical content.

Eqs.~\eqref{eq:RF_time_definition} and \eqref{eq:RF_laplace}
therefore provide both an exact all-orders representation of the
resolvent and the successive finite-order hierarchy used in the
numerical evaluation. Although \(k_{\scriptstyle{F}}(u)\) has an
integrable square-root singularity at the origin,
\[
    |k_{\scriptstyle{F}}(u)|
    =
    C u^{-1/2}\,,
    \qquad
    C=\sqrt{\frac{m}{2 \pi \hbar}}\,,
\]
its successive convolution powers satisfy
\[
    \left|k_{\scriptstyle{F}}^{*n}(t)\right|
    \leq
    C^n
    \frac{\Gamma(1/2)^n}{\Gamma(n/2)}\,
    t^{n/2-1}\,,
    \qquad n\geq 1\,.
\]
The gamma-function growth in the denominator ensures convergence on
every finite time interval. Mathematically, the resulting expansion
is the Neumann series generated by successive substitution in the
second-kind Volterra equation \eqref{eq:Q_convolution}; physically,
its terms are the successive chronological rescattering orders in
Eq.~\eqref{eq:Q_rescattering_hierarchy}.

Finally, substituting Eqs.~\eqref{eq:X-final} and
\eqref{eq:q_resolvent} into Eq.~\eqref{eq:A_b_delta0} reduces the
accumulated contact contribution to
\begin{equation}
\label{eq:Aacc}
    A_b^{(\delta)}(t)
    =
    \frac{i g}{\hbar}
    \left[
        \int_0^{t/2} dw \int_0^{2w} du
        +
        \int_{t/2}^{t}dw
        \int_0^{2(t-w)} du
    \right]
    G_b(u,w)\,,
\end{equation}
where
\begin{equation}
\label{eq:Gb_definition}
    G_b(u,w)
    =
    X_b(u,w)\,
    q\!\left(w-\frac{u}{2}\right)\,.
\end{equation}
Here \(q\) is given by the exact chronological hierarchy
\eqref{eq:Q_rescattering_hierarchy}, or equivalently by the
resolvent form \eqref{eq:q_resolvent}. Eq.~\eqref{eq:Aacc} therefore
retains the complete repeated-contact history while exhibiting
explicitly the convolution structure that generates its successive
orders. This hierarchy, followed by the boundary reduction to the
physical bound-state amplitude, provides the theoretical basis for
the convergence study in Fig.~\ref{fig:bound_state_probability}.

\subsection{Boundary Reduction to the Physical Bound-State Amplitude}
\label{sec:instantaneous_reduction}
%
Having obtained the accumulated bound-state amplitude, we now
recover the physical amplitude through
\begin{equation}
    a_b(t)
    =
    \frac{dA_b(t)}{dt}
    =
    a_b^{(0)}(t)
    +
    a_b^{(\delta)}(t)\,,
    \label{eq:ab_final_decomposition}
\end{equation}
where
\begin{equation}
    a_b^{(0)}(t)
    =
    \frac{dA_b^{(0)}(t)}{dt}
    \;\;\; ;\;\;\;
    a_b^{(\delta)}(t)
    =
    \frac{dA_b^{(\delta)}(t)}{dt}\,.
    \label{eq:ab_components_derivative}
\end{equation}
Applying the Leibniz rule to Eq.~\eqref{eq:Aacc}, as detailed in
Appendix~\ref{app:boundary_reduction}, gives
\begin{equation}
    a_b^{(\delta)}(t)
    =
    \frac{i g}{\hbar}
    \int_0^t du\,
    G_b\!\left(u,t-\frac{u}{2}\right)
    =
    \frac{i g}{\hbar}
    \int_0^t du\,
    X_b\!\left(u,t-\frac{u}{2}\right)\,q(t-u)\,.
    \label{eq:ab_delta_single_integral0}
\end{equation}
Thus, differentiation collapses the two-dimensional accumulated
domain exactly onto its moving boundary \(w = t - u/2\). On this
boundary, the interaction time entering the physical contact
amplitude is \(w - u/2 = t - u\).
Equation~\eqref{eq:ab_delta_single_integral0} therefore contains the
physical chronological amplitude \(q(t-u)\), not a family of
fixed-center-time amplitudes.

Substitution of the all-orders expansion
\eqref{eq:Q_rescattering_hierarchy}, together with \(q(\xi) = e^{i
\Theta(\xi)}\,Q(\xi)\), yields
\begin{equation}
    a_b^{(\delta)}(t)
    =
    \frac{i g}{\hbar}
    \sum_{n=0}^{\infty}
    \left(\frac{i g}{\hbar}\right)^n
    \int_0^t du\,
    X_b\!\left(u,t-\frac{u}{2}\right)\,
    e^{i \Theta(t-u)}\,
    \left(k_{\scriptstyle{F}}^{*n}*Q_0\right)(t-u)\,.
\label{eq:ab_contact_hierarchy}
\end{equation}
Every term in this expression has a direct chronological meaning.
The outer factor \(i g/\hbar\) represents the final contact
encounter followed by projection onto the bound state, while the
index \(n\) counts the preceding internal contact-to-contact
rescattering events. Consequently, \(n = 0\) contains one direct
contact encounter and no internal rescattering, whereas \(n \geq 1\)
describes successively higher rescattering orders. This is precisely
the hierarchy used in the numerical comparison below.

The accumulated-amplitude construction and the convolution
reorganization thus perform distinct but complementary operations.
The endpoint-phase cancellation established in
Eq.~\eqref{eq:chronological_phase_cancellation} converts the
chronological contact series into the relative-time convolution
hierarchy, while differentiation of \(A_b^{(\delta)}(t)\) reduces
the final bound-state projection to the single boundary integral in
Eq.~\eqref{eq:ab_delta_single_integral0}. Neither operation changes
or approximates the physical dynamics.

\subsection{Direct and Repeated-Rescattering Contributions}
%
The resolvent form \eqref{eq:q_resolvent} induces the exact
decomposition
\begin{equation}
    a_b(t)
    =
    a_b^{(0)}(t)
    +
    a_{b,\mbox{\scriptsize{dir}}}^{(\delta)}(t)
    +
    a_{b,\mbox{\scriptsize{resc}}}^{(\delta)}(t)\,.
    \label{eq:Ab_physical_decomposition}
\end{equation}
The contact-free contribution \(a_b^{(0)}(t)\) describes
field-driven propagation followed by projection onto the bound
state, without any subsequent encounter with the contact
interaction. The direct contact contribution follows from the
zeroth-order contact amplitude \(q_0\) and is
\begin{equation}
    a_{b,\mbox{\scriptsize{dir}}}^{(\delta)}(t)
    =
    \frac{i g}{\hbar}
    \int_0^t du\,
    X_b\!\left(u,t-\frac{u}{2}\right)\,
    q_0(t-u)\,.
    \label{eq:ab_dir_final}
\end{equation}
It represents a single encounter with the moving delta-function
contact before the final bound-state projection, with no preceding
contact-to-contact rescattering.

The remaining contribution contains all histories with at least one
internal contact-to-contact rescattering. Using
Eq.~\eqref{eq:q_resolvent} together with \(Q_0(\xi) = e^{-i
\Theta(\xi)}\,q_0(\xi)\), we obtain
\begin{equation}
\begin{aligned}
    a_{b,\mbox{\scriptsize{resc}}}^{(\delta)}(t)
    ={}&
    \frac{i g}{\hbar}
    \int_0^t du\,
    X_b\!\left(u,t-\frac{u}{2}\right)\,
    e^{i\Theta(t-u)}
    \int_0^{t-u} dv\,
    R_{\scriptstyle{F}}(v)\,Q_0(t-u-v)
    \\
    ={}&
    \frac{i g}{\hbar}
    \int_0^t du\,
    X_b\!\left(u,t-\frac{u}{2}\right)
    \int_0^{t-u} dv\,
    R_{\scriptstyle{F}}(v)\,
    \exp\!\left\{
        i\left[
            \Theta(t-u) - \Theta(t-u-v)
        \right]
    \right\}\,
    q_0(t-u-v)\,.
\end{aligned}
\label{eq:ab_mem_final0}
\end{equation}
This form displays explicitly how the endpoint phase of the complete
contact history is restored after the phase factors associated with
all intermediate contact times have canceled. The resolvent
\(R_{\scriptstyle{F}}(v)\) depends only on the relative-time
duration \(v\) and carries no fixed or outer-variable-dependent
center-time label.

Because \(R_{\scriptstyle F}\) is the \(n \geq 1\) part of the
resolvent series in Eq.~\eqref{eq:RF_time_definition},
Eq.~\eqref{eq:ab_mem_final0} collects, or resums, the complete
infinite hierarchy of repeated contact encounters into a single
resolvent contribution. Conversely, expanding \(R_{\scriptstyle F}\)
reproduces term by term the \(n \geq 1\) sector of
Eq.~\eqref{eq:ab_contact_hierarchy}. Consequently, the decomposition
into direct and repeated-rescattering contributions is exact and
introduces no truncation. It is valid for an arbitrary-strength
uniform dc field and at any finite observation time, without a
weak-field expansion, a Born-series truncation, an asymptotic-time
limit, or the explicit construction of the complete KH propagator.

Physically, \(a_{b,\mbox{\scriptsize{resc}}}^{(\delta)}(t)\)
contains histories involving repeated departures from and returns to
the localized interaction region before the final bound-state
projection. It therefore encodes the complete
ionization--recombination history at the contact level. In the
conventional propagator-centered formulation, the same physics
appears as an infinite hierarchy of multiple-scattering terms, as
illustrated explicitly in Appendix~\ref{app:momentum_propagator}. In
the present formulation, this hierarchy is incorporated exactly into
the relative-time resolvent \(R_{\scriptstyle{F}}\) and the
associated rescattering contribution
\(a_{b,\mbox{\scriptsize{resc}}}^{(\delta)}(t)\), without
constructing or truncating its individual terms.

For the present problem, the analytical structure of the exact
construction may be summarized as
\begin{equation}
\begin{aligned}
    K_c(t,\tau)
    &\;\longrightarrow\;
    \text{endpoint-phase factorization}
    \;\longrightarrow\;
    k_{\scriptstyle F}\text{-convolution hierarchy}
    \\
    &\;\longrightarrow\;
    \text{resolvent }\mathcal{R}_{\scriptstyle F}
    =
    \delta + R_{\scriptstyle F}
    \;\longrightarrow\;
    \text{insertion into }A_b^{(\delta)}(t)
    \\
    &\;\longrightarrow\;
    (u,w)\text{-domain organization}
    \;\longrightarrow\;
    a_b^{(\delta)}(t)
    =
    \frac{d A_b^{(\delta)}(t)}{dt}
    \\
    &\;\longrightarrow\;
    a_b(t)
    =
    a_b^{(0)}(t) + a_b^{(\delta)}(t)\,.
\end{aligned}
\label{eq:analytical_chain}
\end{equation}
This sequence shows that the convolution structure is generated by
the complete chronological contact history after its endpoint phases
have been separated. Thus, once again, it is not obtained by
imposing a common fixed center time on successive physical
propagation intervals.

\subsection{Complete Finite-Time Representation and Numerical
Verification} \label{subsec:complete_finite-time}
%
We now collect the explicit ingredients entering the complete
finite-time amplitude. The contact-free contribution can be
evaluated independently of the contact hierarchy. A detailed
derivation is given in Appendix~\ref{app:ab0_derivation}.

Introduce
\begin{equation}
    \eta(t)
    \equiv
    \frac{p_c(t)}{\hbar}\,,
    \label{eq:eta0}
\end{equation}
together with the four complex poles
\begin{equation}
    z_1=\frac{\eta}{2}+i\kappa
    \;\;\; ;\;\;\;
    z_2=\frac{\eta}{2}-i\kappa
    \;\;\; ;\;\;\;
    z_3=-\frac{\eta}{2}+i\kappa
    \;\;\; ;\;\;\;
    z_4=-\frac{\eta}{2}-i\kappa\,,
    \label{eq:ab0_poles}
\end{equation}
and the coefficients
\begin{equation}
    c_j
    =
    \prod_{\ell\neq j}
    \frac{1}{z_j - z_\ell}\,.
    \label{eq:cj_definition}
\end{equation}
The Faddeeva function is defined as \cite{FAD61,DLM26}
\begin{equation}
    w(z)
    =
    e^{-z^2} \operatorname{erfc}(-iz)\,.
    \label{eq:Faddeeva_definition}
\end{equation}
To treat the poles in the upper and lower half-planes of the complex
\(z\)-plane consistently, we introduce
\begin{equation}
    Z_j
    =
    e^{i\pi/4}
    \sqrt{\frac{\hbar t}{2m}}\,z_j
    \;\;\; ;\;\;\;
    \sigma_j
    =
    \operatorname{sgn}\!\left(\operatorname{Im}z_j\right)\,.
    \label{eq:Zj_sigmaj}
\end{equation}
As detailed in Appendix~\ref{app:ab0_derivation}, the sign
prescription \(\sigma_j\) implements the required analytic
continuation for poles in the two half-planes. For \(\eta\neq0\),
the contact-free contribution then takes the closed form
\begin{equation}
    a_b^{(0)}(t)
    =
    2 i \kappa^3
    \exp\!\left(-\frac{i F_0^2 t^3}{24 m \hbar}\right)
    \sum_{j=1}^{4}
    \sigma_j c_j\,
    w\!\left(\sigma_j Z_j\right)\,.
    \label{eq:ab0_Faddeeva}
\end{equation}
In the field-free case \(F_0 = 0\), for which \(\eta = 0\),
Eq.~\eqref{eq:ab0_Faddeeva} is understood through the continuous
limit \(\eta \rightarrow 0\), since pairs of simple poles coalesce
in this limit. The corresponding field-free consistency check and
the verification of the pole prescription are presented in
Appendix~\ref{app:field_free_consistency}.

Eq.~\eqref{eq:ab0_Faddeeva} is a genuine closed-form expression in
terms of a standard special function rather than an unevaluated
momentum integral. Its evaluation requires neither the physical
contact amplitude \(q(t)\) nor the contact resolvent, and it does
not require construction of the complete KH propagator. It gives the
field-driven contact-free contribution directly at the level of the
physical bound-state projection. By gauge equivalence, the same
physical contribution can also be obtained in the scalar- and
vector-potential gauges when the final projection is transformed
consistently. The corresponding scalar-gauge result was obtained in
a different form in Ref.~\cite{ELK88}.

We next examine the direct contact contribution obtained in
Eq.~\eqref{eq:ab_dir_final}. Because \(X_b(u,w)\) and \(q_0(t)\) are
given explicitly by Eqs.~\eqref{eq:X-final} and \eqref{eq:q_0-1},
respectively, Eq.~\eqref{eq:ab_dir_final} provides an exact and
explicitly evaluable one-dimensional time integral. It is therefore
natural to ask whether this remaining integral admits a further
closed-form reduction. The spatial integration has already been
carried out in the projection block \(X_b(u,w)\). On the boundary
\(w = t - u/2\), Eq.~\eqref{eq:Delta_uw} gives
\begin{equation}
    \Delta_t(u)
    \equiv
    \Delta\!\left(u,t - \frac{u}{2}\right)
    =
    \frac{F_0 u}{m}
    \left(t - \frac{u}{2}\right)\,.
    \label{eq:Delta_t_definition}
\end{equation}
Using Eq.~\eqref{eq:B_sigma_general}, the parameter in the
corresponding half-line Gaussian--\(\operatorname{erfc}\) block then
reduces to
\begin{equation}
    B_\sigma\!\left(u;\Delta_t(u),F_0 t\right)
    =
    -\kappa
    +
    \frac{i \sigma F_0 t}{\hbar}
    -
    2 i \sigma\frac{m}{2 \hbar u} \Delta_t(u)
    =
    -\kappa
    +
    \frac{i \sigma F_0 u}{2 \hbar}\,.
\label{eq:B_sigma_direct_simplified}
\end{equation}
Similarly, on writing \(v = t - u\), the parameter entering
\(q_0(v)\) is
\begin{equation}
    B_\rho\!\left(v;-x_c(v),0\right)
    =
    -\kappa
    +
    \frac{i \rho F_0 v}{2 \hbar}\,.
    \label{eq:B_rho_q0_simplified}
\end{equation}
Although these parameters simplify, the product of the two half-line
Gaussian--\(\operatorname{erfc}\) blocks retains both square-root
and three-half-power time dependences. For a general nonzero dc
field, no reduction of the remaining finite-interval integral to a
single standard Faddeeva expression is apparent.
Eq.~\eqref{eq:ab_dir_final} is therefore retained as the final exact
one-dimensional representation of the direct contact contribution.

Combining Eqs.~\eqref{eq:ab_dir_final}, \eqref{eq:ab_mem_final0},
and \eqref{eq:ab0_Faddeeva}, we obtain the complete finite-time
amplitude in the exact resolvent form
\begin{subequations}
\label{eq:ab_explicit_final}
\begin{align}
    a_b(t)
    ={}&
    a_b^{(0)}(t)
    +
    \frac{i g}{\hbar}
    \int_0^t du\,
    X_b\!\left(u,t-\frac{u}{2}\right)\,
    q_0(t-u)
    \nonumber\\
    &+
    \frac{i g}{\hbar}
    \int_0^t du\,
    X_b\!\left(u,t-\frac{u}{2}\right)
    \int_0^{t-u} dv\,
    R_{\scriptstyle F}(v)\,
    \exp\!\left\{
        i \left[
            \Theta(t-u) - \Theta(t-u-v)
        \right]
    \right\}\,
    q_0(t-u-v)\,.
    \label{eq:ab_explicit_resolvent}
\end{align}
Equivalently, its chronological rescattering hierarchy is
\begin{equation}
    a_b(t)
    =
    a_b^{(0)}(t)
    +
    \frac{i g}{\hbar}
    \sum_{n=0}^{\infty}
    \left(\frac{i g}{\hbar}\right)^n
    \int_0^t du\,
    X_b\!\left(u,t-\frac{u}{2}\right)
    \int_0^{t-u} dv\,
    k_{\scriptstyle F}^{*n}(v)\,
    e^{i [
            \Theta(t-u) - \Theta(t-u-v)
        ]}\,
    q_0(t-u-v)\,.
    \label{eq:ab_explicit_hierarchy}
\end{equation}
\end{subequations}
Eq.~\eqref{eq:ab_explicit_resolvent} is the exact resolvent form, in
which all repeated-rescattering orders are collected into
\(R_{\scriptstyle F}\), while Eq.~\eqref{eq:ab_explicit_hierarchy}
displays those orders individually and, upon truncation of the sum,
yields the successive finite-order approximations used in the
numerical calculation. All spatial integrations have already been
carried out analytically. The direct contact contribution therefore
requires only a one-dimensional time integral, while the
repeated-rescattering contribution is represented by a
two-dimensional time integral involving the exact resolvent, or
equivalently by the \(n \geq 1\) sector of the convergent
finite-time convolution hierarchy.

\begin{figure}[t]
\centering
\includegraphics[width=\textwidth]{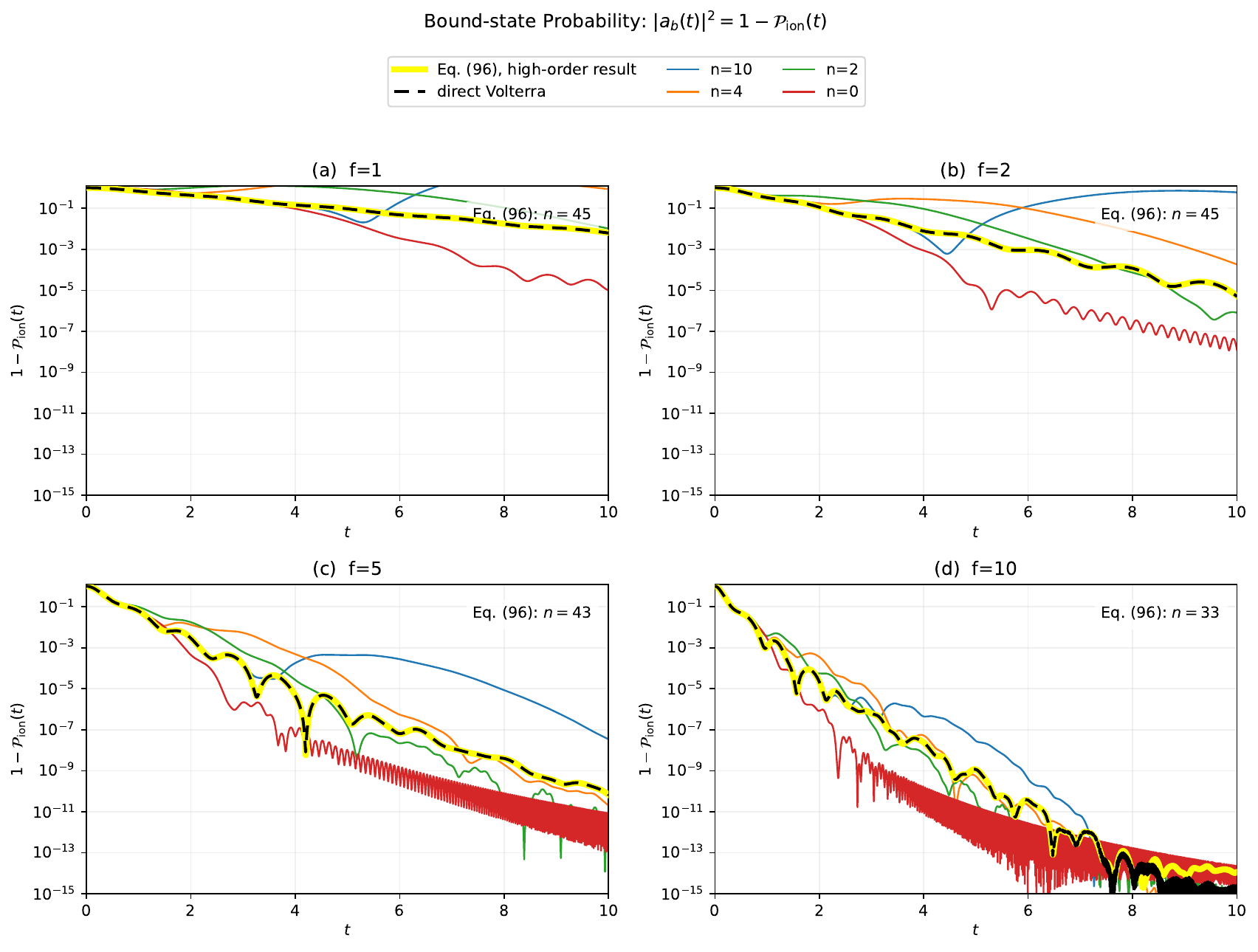}
\caption{Bound-state survival probability \(1 -
\mathcal{P}_{\mbox{\scriptsize{ion}}}(t) =\lvert a_b(t)\rvert^2\)
versus time \(t\) for \(f = 1, 2, 5,\) and \(10\). The curves
labeled by \(n\) are successive truncations of the exact
chronological rescattering hierarchy in
Eq.~\eqref{eq:ab_explicit_hierarchy}, where \(n\) counts internal
contact-to-contact rescattering events. Thus, \(n = 0\) contains the
contact-free contribution and the single direct contact
contribution, but no internal contact-to-contact rescattering. For
the numerical implementation and tolerance used here, the lowest
orders displayed as converged are \(n = 45, 45, 43,\) and \(33\) for
\(f = 1, 2, 5,\) and \(10\), respectively. These thresholds are
tolerance- and implementation-dependent. The black dashed curves are
obtained independently by numerically solving the original physical
Volterra integral equation \eqref{eq:kh_contact_equation} directly
for \(q(t)\) and substituting the result into
Eq.~\eqref{eq:kh_direct_amplitude}. The finite-order hierarchy and
the direct numerical Volterra solution are nearly indistinguishable
over the displayed range. At very late times for \(f = 10\), where
the survival probability decreases through the range
\(10^{-12}\)--\(10^{-15}\), the small residual deviation is
attributable to the time discretization used in the direct Volterra
calculation rather than to incomplete convergence of the
rescattering hierarchy.} \label{fig:bound_state_probability}
\end{figure}

The ionization probability \({\mathcal
P}_{\mbox{\scriptsize{ion}}}(t)\) introduced in Sec.~\ref{sec:intro}
follows directly from the equivalent finite-time representations
collected in Eq.~\eqref{eq:ab_explicit_final}. For the numerical
comparison, we use the dimensionless field strength \cite{ELK88}
\begin{equation}
    f
    =
    \frac{m F_0}{\hbar^2 \kappa^3}\,,
    \label{eq:dimensionless_field_strength}
\end{equation}
which measures the applied dc field relative to the characteristic
field scale of the bound state. The present result is
nonperturbative in \(f\). In contrast to earlier weak-field
treatments, such as our treatment in Ref.~\cite{KIM07}, which
assumes \(f \ll 1\), the present formulation remains valid for
arbitrary dc-field strength and arbitrary finite observation time,
without a weak-field expansion, a rescattering truncation, or an
asymptotic-time approximation.
Fig.~\ref{fig:bound_state_probability} shows the systematic
convergence of the finite-order hierarchy toward the result obtained
by directly solving the original physical Volterra integral
equation. Their agreement provides an independent internal check of
the endpoint-phase cancellation, the convolution reorganization, and
the rescattering-order counting.

In connection with this comparison, we note a possible discrepancy
in the rescattering expression reported in Ref.~\cite{ELK88}. The
Volkov propagator in Eq.~(2.8) of that work contains, through the
phase defined in its Eq.~(2.10), the classical-action difference
\(-[S_{\rm cl}(t)-S_{\rm cl}(t')]\), which corresponds in the
notation of the present work to \(-[s_c(t) - s_c(t')]\), with
\(s_c(t)\) given in Eq.~\eqref{eq:action_dc}. However, this term
appears to be lost in the reduction from its Eq.~(2.23) to its
Eq.~(2.24): the phase \(\mu(t,t')\) printed in its Eq.~(2.25)
contains the remaining two terms but not the classical-action
difference. Consequently, the corresponding phase factor is also
absent from the constant-field rescattering expression in its
Eq.~(3.6). This action term is dynamically essential. In the
contact-to-contact Volkov kernel for a constant field, it combines
with the free-propagator displacement phase to produce the cubic
exponential factor in Eq.~\eqref{eq:scalar_cubic_phase}, which is
precisely the relative-time factor \(\exp(-i \gamma u^3)\) in
Eq.~\eqref{eq:kF_definition}. This apparent omission may therefore
be related to the quantitative deviations between the published
strong-field curves in Ref.~\cite{ELK88} for \(f = 5\) and \(10\)
and the present results. Since the physical bound-state probability
is representation independent, these deviations cannot be attributed
to the choice of representation alone. Independently, the agreement
between the convolution hierarchy and the direct numerical solution
of the original physical Volterra integral equation in
Fig.~\ref{fig:bound_state_probability} provides an important
internal check of the phase structure and finite-time dynamics
obtained here.

The field-free limit supplies a further stringent consistency test.
As shown in Appendix~\ref{app:field_free_consistency}, setting \(F_0
= 0\) gives \(\Theta = 0\) and reduces \(k_{\scriptstyle F}\) and
\(R_{\scriptstyle F}\) to their field-free convolution kernels. The
contact-free and contact contributions then combine to recover
\(a_b(t) = e^{i \Omega t}\), where \(\Omega = \hbar \kappa^2/(2
m)\), and hence \(|a_b(t)|^2 = 1\). The same calculation also
verifies the \(\sigma_j\)-dependent Faddeeva pole prescription.

\subsection{Computational Significance of the KH Result}
\label{subsec:computational_significance}
%
The principal computational result is the explicit finite-time
physical amplitude in Eq.~\eqref{eq:ab_explicit_final}, obtained
without constructing the complete chronologically ordered KH
propagator. The KH transformation converts the external-force
problem into free propagation between encounters with a moving
contact. For a dc trajectory, the resulting contact kernel admits
the exact endpoint-phase factorization \eqref{eq:Kc_factorization}.
The internal endpoint phases then cancel throughout every
chronological history, leaving the single cubic-phase kernel
\(k_{\scriptstyle{F}}\) and its ordinary convolution powers.

The accumulated amplitude supplies the second reduction. It embeds
the final projection into the exact two-time domain, and its
derivative collapses that domain to the one-dimensional boundary
integral in Eq.~\eqref{eq:ab_delta_single_integral0}. Thus, the
contact dynamics is handled by the exact convolution resolvent
\(R_{\scriptstyle F}\), while the bound-state projection is handled
by the explicitly evaluated block \(X_b\). These two reductions
together produce the resolvent and hierarchy forms in
Eq.~\eqref{eq:ab_explicit_final}.

An equally explicit finite-time amplitude is not readily obtained by
constructing the complete propagator first in the scalar- or
vector-potential gauge. The distinction lies not in the physical
dynamics, which is gauge equivalent, but in the analytical object
chosen from the outset. By organizing the calculation around the
desired bound-state amplitude, the present formulation avoids the
larger task of constructing a propagator containing all spatial
matrix elements before extracting a single physical projection.

To the best of our knowledge, an exact finite-time representation of
the instantaneous ionization amplitude comparable to
Eq.~\eqref{eq:ab_explicit_final} has not previously been reported
for the uniformly driven attractive delta well. The result is
nonperturbative in the dc-field strength and contains the complete
contact-rescattering hierarchy. Its numerical evaluation may use
either the resolvent form or controlled successive convolution
orders, with the direct physical Volterra solution providing an
independent verification.

\section{Conclusions}
\label{sec:conclusions}
%
We have developed an exact amplitude-centered formulation of
finite-time ionization from a one-dimensional attractive
delta-function well subjected to an arbitrary-strength uniform dc
electric field. The formulation bypasses the prior construction of
the complete time-dependent propagator or wavefunction by combining
the Kramers--Henneberger moving-contact representation, the
accumulated bound-state amplitude, and a relative- and center-time
reorganization, while retaining the complete chronological contact
dynamics.

The central analytical result is the exact factorization of the dc
moving-contact kernel into two endpoint phases and a single
relative-time cubic-phase kernel. In a complete chronological
rescattering history, the phases at all intermediate contact times
cancel pairwise. The remaining factors depend only on the durations
of successive propagation intervals and therefore form an ordinary
Volterra convolution hierarchy generated by \(k_{\scriptstyle
F}(u)\). Equivalently, the rephased physical contact amplitude
\(Q(t)=e^{-i\Theta(t)}q(t)\) satisfies an exact convolution equation
whose all-orders solution is encoded in the resolvent
\(\mathcal{R}_{\scriptstyle F}\). This mechanism does not assume a
common fixed center time for successive physical intervals and
introduces no approximation to the original Volterra dynamics.

The accumulated amplitude plays a complementary role. After the
physical contact hierarchy is inserted into the bound-state
projection, differentiation with respect to the observation time
collapses the two-dimensional \((u,w)\) domain onto its boundary.
The physical contact contribution is thereby reduced exactly to a
single outer relative-time integral. Its zeroth-order contact sector
gives the direct contribution, while the convolution resolvent gives
the complete repeated-rescattering contribution. Combining these two
contact contributions with the contact-free contribution
\(a_b^{(0)}(t)\), obtained independently from the field-driven
propagation without a contact encounter, yields the complete
finite-time bound-state amplitude. This amplitude is expressed
equivalently in the exact resolvent form
\eqref{eq:ab_explicit_resolvent} and the chronological hierarchy
form \eqref{eq:ab_explicit_hierarchy}.

All spatial integrations entering the final amplitude have been
performed analytically. The contact-free contribution is obtained in
closed form in terms of the Faddeeva function, the direct contact
contribution is an explicitly evaluable one-dimensional time
integral, and the complete repeated-rescattering contribution is an
explicitly evaluable two-dimensional time integral involving the
exact, center-time-independent resolvent \(R_{\scriptstyle F}\).
This reduction contrasts with the conventional propagator-centered
hierarchy illustrated in Appendix~\ref{app:momentum_propagator}, in
which successive scattering orders retain nested momentum and time
integrations before the final bound-state projection is performed.
In contrast to the weak-field regime \(f \ll 1\) considered in our
earlier treatment in Ref.~\cite{KIM07}, the present formulation is
valid at arbitrary finite observation time and arbitrary dc-field
strength and requires no weak-field expansion, rescattering
truncation, or asymptotic-time approximation. The resulting
amplitude directly determines the finite-time bound-state survival
probability through \(\lvert a_b(t)\rvert^2\).

Two independent checks support the result. First, the exact
field-free limit recovers \(a_b(t)=e^{i\Omega t}\) and
\(|a_b(t)|^2=1\), including the required cancellation between the
contact-free and contact contributions and the exact Faddeeva pole
prescription. Second, successive truncations of the chronological
convolution hierarchy converge to the result obtained by solving the
original physical Volterra equation directly. This agreement
independently verifies the cubic-phase kernel, the rescattering
counting, and the final boundary-reduced amplitude.

The principal methodological conclusion is that, within the present
KH formulation, the physical finite-time amplitude can be
constructed from the contact dynamics without explicitly
reconstructing the complete chronological propagator or
wavefunction. This conclusion does not imply that the physical
amplitude is available only in the KH representation; by gauge
equivalence, the same amplitude may also be formulated in the
scalar- and vector-potential gauges. The particular advantage of the
KH representation is structural: it expresses the driven dynamics as
free propagation between encounters with a moving contact and
thereby makes the endpoint-phase separation, the relative-time
convolution hierarchy, and the boundary reduction of the accumulated
amplitude especially transparent. Beyond the present model, these
complementary reductions suggest a broader amplitude-centered
analytical strategy for nonperturbative driven quantum dynamics
whenever the interaction kernel admits a corresponding separation
into endpoint factors and a relative-time propagation kernel.
Periodically driven systems provide a natural next setting, in which
the remaining center-time dependence may acquire a Floquet structure
rather than cancel completely.

\section*{AUTHOR DECLARATIONS}
%
\subsection*{Conflict of Interest}
The authors have no conflicts to disclose.

\subsection*{Author Contributions}
I. Kim: Conceptualization (lead); Methodology (lead); Formal
analysis (lead); Software; Validation; Visualization; Writing --
original draft; Writing -- review \& editing.

G. J. Iafrate: Conceptualization (supporting); Methodology
(supporting); Writing -- review \& editing.

\section*{DATA AVAILABILITY}
%
The data that support the findings of this study are available from
the corresponding author upon reasonable request.

\appendix
%
%
\section{Field-Free Attractive Delta-Function Potential}
\label{app:field_free_case}
%
We now verify that the present formulation developed in
Sec.~\ref{sec:kh_formulation} recovers the exact field-free
solution. Setting \(F_0 = 0\), the Lippmann--Schwinger integral
equation, Eq.~\eqref{eq:scalar_lippmann_schwinger_propagator},
reduces to the standard field-free integral equation \cite{ELB88}.
Evaluating the corresponding field-free wavefunction equation,
Eq.~\eqref{eq:scalar_wavefunction_from_full_propagator}, obtained by
setting \(F_0 = 0\), at the contact point \(x = 0\), and then
applying the Laplace transform \(\mathcal{L}\), together with the
convolution theorem, yields
\begin{equation}
\label{eq:laplace_psi_x0_a.c}
    \tilde{\psi}_0(0,s)
    =
    \frac{\tilde{\phi}_0(0,s)}
    {1
    -
    i g
    \sqrt{m/(2 i \hbar^3\,s)}}\,,
\end{equation}
where
\[
\tilde{\psi}_0(x,s) = \mathcal{L}\{\psi_0(x,t)\}\;\;\; ;\;\;\;
\tilde{\phi}_0(x,s) = \mathcal{L}\{\phi_0(x,t)\}\,,
\]
with the homogeneous contribution \(\phi_0(x,t)\) given by
Eq.~\eqref{eq:scalar_homo_wavefunction_from_full_propagator} for
\(F_0 = 0\). Furthermore,
\[
\tilde{\phi}_0(0,s) = \frac{\sqrt{\kappa}}{\sqrt{s}\; (\sqrt{s} +
\sqrt{-i E_b/\hbar})}\,,
\]
which follows from
\begin{equation}\label{eq:laplace_psi_x0_f0}
    \phi_0(0,t)
    =
    \sqrt{\kappa}\,
    e^{-i E_b t/\hbar}
    \,
    \mathrm{erfc}
    \!\left(
    \sqrt{-\frac{i E_b t}{\hbar}}
    \right)\,,
\end{equation}
expressed in terms of the complementary error function.

Applying the inverse Laplace transform to
Eq.~\eqref{eq:laplace_psi_x0_a.c} using
Eq.~\eqref{eq:laplace_psi_x0_f0} yields \cite{ROB66}
\[
\psi_0(0,t) = \sqrt{\kappa}\, e^{-i E_b t/\hbar}\,.
\]
Substituting this result into
Eq.~\eqref{eq:scalar_wavefunction_from_full_propagator}, together
with Eq.~\eqref{eq:scalar_lippmann_schwinger_propagator} for \(F_0 =
0\), gives
\begin{eqnarray}\label{eq:wave_function_without_field}
    \psi_0(x,t) &=& \phi_0(x,t) + \sqrt{\kappa} \left[M\!\left(|x|; i \kappa; \frac{\hbar t}{m}\right) - M\!\left(|x|; -i \kappa; \frac{\hbar t}{m}\right)\right]\nonumber\\
    &=& \sqrt{\kappa} \left[M\!\left(|x|; i \kappa; \frac{\hbar t}{m}\right) + M\!\left(-|x|; -i \kappa; \frac{\hbar t}{m}\right)\right]\, =\,
    \psi_b(x)\,e^{-i E_b t/\hbar}\,,
\end{eqnarray}
where the Moshinsky function is defined by
\[
M(x;k;t) = \frac{1}{2}\,e^{i(k x - k^2
t/2)}\,\mathrm{erfc}\!\left(\frac{x - k t}{\sqrt{2it}}\right)\,,
\]
following Refs.~\cite{MOS52,MOS76}. Thus, the present field-driven
formulation exactly reproduces the field-free bound-state evolution
in the limit \(F_0 \rightarrow 0\), providing an independent
consistency check of the field-driven formulation developed in the
main text. Along the same lines, the exact field-free propagator
\(\mathcal{K}_0(x,t|x',0)\) was also derived in \cite{ELB88}.

\section{Conventional Propagator Formulation in Momentum Space}
\label{app:momentum_propagator}
%
Although the present formulation avoids the explicit construction of
the chronological propagator, it is instructive to examine the
corresponding momentum-space formulation obtained within the
conventional propagator-centered approach. The purpose of this
appendix is not to develop an alternative formulation, but to
illustrate the rapid growth of analytical complexity inherent in the
conventional propagator formulation. As shown below, repeated
scattering by the delta-function potential generates an infinite
hierarchy of multiple ionization--recombination processes, making
the explicit construction of the propagator substantially more
demanding than the present amplitude-centered approach. To
illustrate this point, we briefly summarize the conventional
momentum-space formulation in the vector-potential gauge. The
derivation follows the standard propagator-based approach and leads
directly to an infinite multiple-scattering expansion. The resulting
expressions make transparent how the analytical complexity grows
rapidly, even for the one-dimensional delta-function potential.

We begin by considering the momentum representation of the
propagator $\mathcal{K}_{\mbox{\scriptsize{s}}}(x,t|x',0)$,
\begin{equation}\label{eq:momentum_representaion_k_1}
    \<p|\hat{U}_{\mbox{\scriptsize{s}}}(t)|p'\> =
    U_{\mbox{\scriptsize{s}}}(p,t|p',0) = \int_{-\infty}^{\infty} dx\, \int_{-\infty}^{\infty} dx'\,
    \<p|x\>\, \mathcal{K}_{\mbox{\scriptsize{s}}}(x,t|x',0)\, \<x'|p'\>\,,\n
\end{equation}
where $\hat{U}_{\mbox{\scriptsize{s}}}(t) = e^{-\frac{i}{\hbar}
\hat{H}_{\mbox{\tiny{s}}} t}$ and $\<p|x\> = e^{-\frac{i}{\hbar} x
p}/\sqrt{2 \pi \hbar}$. By setting \(x = 0\) in
Eq.~\eqref{eq:scalar_lippmann_schwinger_propagator}, iteratively
substituting the integrand
\(\mathcal{K}_{\mbox{\scriptsize{s}}}(0,\tau|x',0)\) by the
right-hand side of the same equation, and subsequently performing
the Fourier transform term by term, we can obtain an explicit
expression for \(U_{\mbox{\scriptsize{s}}}(p,t|p',0)\). After a
lengthy calculation that takes into account
$\<p|\hat{U}_{\mbox{\scriptsize{v}},\scriptstyle{F}}(t)|p'\>$
together with Eq.~\eqref{eq:scalar_vector_state_relation}, we arrive
at
\begin{equation}\label{eq:momentum_representaion_k_2}
    U_{\mbox{\scriptsize{s}}}(p,t|p',0) =
    \mathcal{A}_{{\mbox{\scriptsize{s}}}\leftarrow{\mbox{\scriptsize{v}}},\scriptstyle{F}}(p,p',t)\;
    U_{\mbox{\scriptsize{v}},\scriptstyle{F}}(p-p_c(t),t)\,,
\end{equation}
where $U_{\mbox{\scriptsize{v}},\scriptstyle{F}}(p-p_c(t),t) \equiv
U_{\mbox{\scriptsize{v}},\scriptstyle{F}}(p-p_c(t);t,0)$ in
Eq.~\eqref{eq:vector_field_evolution_factor}, and
\begin{eqnarray}\label{eq:momentum_representaion_k_3}
    \mathcal{A}_{{\mbox{\scriptsize{s}}}\leftarrow{\mbox{\scriptsize{v}}},\scriptstyle{F}}(p,p',t)\; &\equiv& \delta(p-p_c(t)-p')
    + \int_0^t d\tau_1\, U_{\mbox{\scriptsize{v}},\scriptstyle{F}}^{\ast}(p-p_c(t),\tau_1) \times\n\\
    && \sum_{k=1}^{\infty} \left(\frac{i g}{2 \pi \hbar^2}\right)^{k}\,
    \left[\prod_{\ell=1}^{k-1} \int_0^{\tau_{\ell}} d\tau_{\ell+1}
    \int_{-\infty}^{\infty} dp_{\ell}\,
    U_{\mbox{\scriptsize{v}},\scriptstyle{F}}(p_{\ell},\tau_{\ell})\,
    U_{\mbox{\scriptsize{v}},\scriptstyle{F}}^{\ast}(p_{\ell},\tau_{\ell+1})\right]
    U_{\mbox{\scriptsize{v}},\scriptstyle{F}}(p',\tau_{k})\,.\n
\end{eqnarray}
Here, the summation index \(k\) counts the number of interactions
with the delta-function well during propagation from the initial
momentum \(p'\) to the final momentum \(p\). The Born term (\(k=1\))
can be expressed in closed form in terms of the Moshinsky function
introduced in Appendix~\ref{app:field_free_case}:
\begin{equation}\label{eq:born_term_momentum_rep}
    \mathcal{A}_{{\mbox{\scriptsize{s}}}\leftarrow{\mbox{\scriptsize{v}}},\scriptstyle{F}}^{(\mbox{\scriptsize{B}})}(p,p',t) = \frac{i
    g}{\hbar} \sqrt{\frac{i m} {2 \pi \hbar F_0\,D(t)}}\,
    \left[M\!\left(0;k_1(t);T_{\scriptstyle{F}}(t)\right) - \exp\!\left[\frac{i D(t)\,t\,(S(t) + F_0 t)}{2 \hbar m}\right]
    M\!\left(0;k_2(t);T_{\scriptstyle{F}}(t)\right)\right]\n\,,
\end{equation}
where $D(t)=p-p'-p_c(t)$ and $S(t)=p+p'-p_c(t)$, with
\[T_{\scriptstyle{F}}(t) =
\frac{\hbar m}{F_0\,D(t)}\;\;\; ;\;\;\; k_1(t) =
-\frac{D(t)\,S(t)}{2 \hbar m}\;\;\; ;\;\;\; k_2(t) =
-\frac{D(t)\,(S(t) + 2F_0 t)}{2 \hbar m}\,.
\]
This scalar-gauge momentum representation, constructed from the
vector-gauge propagators
\(\{U_{\mbox{\scriptsize{v}},\scriptstyle{F}}(p,t)\}\), clearly
demonstrates the complexity of the bound--continuum dynamics
involving infinitely many continuum--continuum rescattering
processes.

The infinite multiple-scattering expansion derived above illustrates
the rapidly increasing analytical complexity inherent in the
conventional propagator-centered formulation. Although formally
exact, this approach requires the systematic summation of infinitely
many ionization--recombination histories before the physical
bound-state amplitude can be extracted. By contrast, the
amplitude-centered formulation developed in the main text treats the
bound-state amplitude itself as the fundamental analytical object.
The resulting temporal reorganization then leads directly to an
exact finite-time solution without first constructing the complete
chronological propagator.

\section{Transformation to Relative- and Center-Time Variables}
\label{app:uw_derivation}
%
For completeness, we provide the detailed derivation of
Eq.~\eqref{eq:A_b_delta0}. Substituting
Eqs.~\eqref{eq:kh_direct_amplitude} and
\eqref{eq:kh_bound_contact_kernel} into
Eq.~\eqref{eq:Ab_delta_definition}, we obtain
\begin{equation}
    A_b^{(\delta)}(t)
    =
    \frac{i g}{\hbar}
    \int_0^t ds
    \int_0^s d\tau
    \int_{-\infty}^{\infty} dx\;
    \chi_b^*\!\left(
        x + x_c(s)
    \right)\,
    K_0\!\left(
        x,s
        \middle|
        -x_c(\tau),\tau
    \right)\,
    q(\tau)\,.
    \label{eq:Ab_delta_st}
\end{equation}
Using the inverse transformation in Eq.~\eqref{eq:uw_inverse} and
the transformed domain in Eq.~\eqref{eq:uw_domain_integral}, this
expression becomes
\begin{equation}
    A_b^{(\delta)}(t)
    =
    \frac{i g}{\hbar}
    \int_0^t du
    \int_{u/2}^{t-u/2} dw
    \int dx\;
    \chi_b^*\!\left(
        x + x_c\!\left(w + \frac{u}{2}\right)
    \right)\,
    K_0\!\left(
        x,w + \frac{u}{2}
        \middle|
        -x_c\!\left(w - \frac{u}{2}\right),
        w - \frac{u}{2}
    \right)\,
    q\!\left(w - \frac{u}{2}\right)\,.
\label{eq:Ab_delta_uw}
\end{equation}
Since \(s - \tau = u\), the free propagator appearing in
Eq.~\eqref{eq:Ab_delta_uw} takes the explicit form
\begin{equation}
    K_0\!\left(
        x,w + \frac{u}{2}
        \middle|
        -x_c\!\left(w - \frac{u}{2}\right),
        w - \frac{u}{2}
    \right)
    =
    \sqrt{\frac{m}{2 \pi i \hbar u}}\,
    \exp\!\left[
        \frac{i m}{2 \hbar u}
        \left(
            x + x_c\!\left(w - \frac{u}{2}\right)
        \right)^2
    \right]\,.
\label{eq:K0_uw_explicit}
\end{equation}
Finally, we interchange the order of the \(u\) and \(w\)
integrations according to Eq.~\eqref{eq:wu_domain_integral0}. Using
Eq.~\eqref{eq:bound_state_in_KH} together with
Eq.~\eqref{eq:K0_uw_explicit}, we then obtain
Eqs.~\eqref{eq:A_b_delta0} and \eqref{eq:F_uw}.

\section{Explicit Evaluation of the Homogeneous Contact Amplitude}
\label{app:q0_derivation}
%
Here we derive the explicit expression for the homogeneous
(contact-free) contact amplitude \(q_0(t)\) used in
Eq.~\eqref{eq:q_0-1}. From the definition of the homogeneous KH
wavefunction, together with Eqs.~\eqref{eq:bound_state} and
\eqref{eq:scalar_dc_field_propagator} evaluated at \(F_0=0\) and
\(x=-x_c(t)\),
\begin{eqnarray}
    q_0(t)
    &=&
    \phi_{\mbox{\scriptsize KH}}(-x_c(t),t)
    =
    \int_{-\infty}^{\infty} dx'\,
    K_0(-x_c(t),t|x',0)\,
    \psi_b(x')\n\\
    &=&
    \sqrt{\frac{\kappa\,m}{2 \pi i \hbar t}}
    \int_{-\infty}^{\infty} dx'\,
    e^{-\kappa |x'|}
    \exp\!\left[
        \frac{i m}{2 \hbar t}
        \left[x' + x_c(t)\right]^2
    \right]\,.
    \label{eq:q_0_appendix}
\end{eqnarray}
To evaluate this integral, split the real line into the two
half-lines and write \(x' = \sigma r\), with \(r \ge 0\) and
\(\sigma=\pm1\). Then, Eq.~\eqref{eq:q_0_appendix} becomes
\begin{equation}\label{eq:app_int1}
    q_0(t)
    =
    \sqrt{\frac{\kappa\,m}{2 \pi i \hbar t}}
    \sum_{\sigma = \pm 1}
    \int_0^\infty dr\,
    e^{-\kappa r}\,
    \exp\!\left[
        \frac{i m}{2 \hbar t}
        \left[\sigma r + x_c(t)\right]^2
    \right]\,.
\end{equation}
Introducing \(\Delta = -x_c(t)\), the exponent can be rewritten as
\begin{equation}
\begin{aligned}
    \frac{i m}{2 \hbar t}
    \left[\sigma r+x_c(t)\right]^2 - \kappa r
    &=
    \frac{i m}{2 \hbar t}
    \left(r^2 + \Delta^2\right)
    +
    B_\sigma(t;\Delta,0)\,r
\end{aligned}
\end{equation}
with
\begin{equation}
    B_\sigma(t;\Delta,0)
    =
    -\kappa
    -
    \frac{i \sigma m}{\hbar t} \Delta\,.
\end{equation}
Accordingly, define the integral in Eq.~\eqref{eq:app_int1} as
\begin{equation}\label{eq:I_sigma_app}
    I_\sigma(t;\Delta,0)
    \equiv
    e^{i m\,\Delta^2/(2 \hbar t)}
    \int_0^\infty dr\,
    \exp\!\left[
        \frac{i m}{2 \hbar t} r^2
        +
        B_\sigma(t;\Delta,0)\,r
    \right]\,.
\end{equation}
Using the standard half-line Gaussian integral, understood by
analytic continuation for the oscillatory Gaussian, this yields
Eq.~\eqref{eq:I_sigma_q0} with \(p = 0\). For the dc trajectory with
\(x_c(t) = F_0\,t^2/(2 m)\), the parameter entering
Eq.~\eqref{eq:I_sigma_app} reduces to
\begin{equation}
    B_\sigma\!\left(t;-x_c(t),0\right)
    =
    -\kappa
    +
    \frac{i \sigma F_0 t}{2 \hbar}\,.
\end{equation}
Therefore, Eq.~\eqref{eq:q_0-1} follows.

\section{Boundary Reduction of the Accumulated Contact Amplitude}
\label{app:boundary_reduction}
%
Here we provide the detailed derivation of the boundary reduction
used in Sec.~\ref{sec:instantaneous_reduction}. From
Eq.~\eqref{eq:Aacc}, we write
\begin{equation}
    A_b^{(\delta)}(t)
    =
    \frac{i g}{\hbar}
    \left[
        A_1(t) + A_2(t)
    \right]\,,
    \label{eq:app_Ab_A1A2}
\end{equation}
where
\begin{subequations}
\begin{equation}
    A_1(t)
    =
    \int_0^{t/2} dw
    \int_0^{2w} du\,
    G_b(u,w)
    \label{eq:app_A1}
\end{equation}
and
\begin{equation}
    A_2(t)
    =
    \int_{t/2}^{t} dw
    \int_0^{2 (t-w)} du\,
    G_b(u,w)\,.
    \label{eq:app_A2}
\end{equation}
\end{subequations}

For the first sector, the dependence on \(t\) enters only through
the upper limit of the \(w\)-integration. The Leibniz rule therefore
gives
\begin{equation}
    \frac{d A_1(t)}{dt}
    =
    \frac{1}{2}
    \int_0^t du\,
    G_b\!\left(u,\frac{t}{2}\right)\,.
    \label{eq:app_dA1}
\end{equation}
For the second sector, define
\begin{equation}
    H(t,w)
    \equiv
    \int_0^{2 (t-w)} du\,
    G_b(u,w)\,,
    \label{eq:app_H}
\end{equation}
so that
\begin{equation}
    A_2(t)
    =
    \int_{t/2}^{t} dw\, H(t,w)\,.
    \label{eq:app_A2_H}
\end{equation}
Differentiation gives
\begin{equation}
\begin{aligned}
    \frac{d A_2(t)}{dt}
    ={}&
    H(t,t)
    -
    \frac{1}{2}
    H\!\left(t,\frac{t}{2}\right)
    +
    \int_{t/2}^{t} dw\,
    \frac{\partial H(t,w)}{\partial t}\,.
\end{aligned}
\label{eq:app_dA2_general}
\end{equation}
Since the upper limit in Eq.~\eqref{eq:app_H} vanishes at \(w = t\),
it follows that \(H(t,t) = 0\), while
\begin{equation}
    H\!\left(t,\frac{t}{2}\right)
    =
    \int_0^t du\,
    G_b\!\left(u,\frac{t}{2}\right)\,.
    \label{eq:app_Hhalf}
\end{equation}
Furthermore, because \(G_b(u,w)\) has no explicit dependence on the
observation time \(t\) when \(u\) and \(w\) are held fixed,
differentiation of Eq.~\eqref{eq:app_H} yields
\begin{equation}
    \frac{\partial H(t,w)}{\partial t}
    =
    2\,G_b\!\left(2 (t-w),w\right)\,.
    \label{eq:app_dHdt}
\end{equation}
Consequently,
\begin{equation}
    \frac{d A_2(t)}{dt}
    =
    2 \int_{t/2}^{t} dw\,
    G_b\!\left(2 (t-w),w\right)
    -
    \frac{1}{2}
    \int_0^t du\,
    G_b\!\left(u,\frac{t}{2}\right)\,.
\label{eq:app_dA2}
\end{equation}

Combining this result with Eq.~\eqref{eq:app_dA1}, the contributions
from the common boundary \(w = t/2\) cancel exactly. Introducing \(u
= 2 (t - w)\), with \(dw = -du/2\), and using
Eq.~\eqref{eq:app_Ab_A1A2}, we recover
Eq.~\eqref{eq:ab_delta_single_integral0}.

\section{Derivation of the Contact-Free Bound-State Amplitude}
\label{app:ab0_derivation}
%
In this Appendix, we derive the closed-form expression for the
contact-free contribution \(a_b^{(0)}(t)\) used in the main text.
Substituting Eqs.~\eqref{eq:bound_state_in_KH} and
\eqref{eq:kh_homogeneous_wavefunction} into
Eq.~\eqref{eq:kh_direct_amplitude}, together with the momentum
representation of the free propagator,
\begin{equation}
    K_0(x,t|x',0)
    =
    \frac{1}{2 \pi \hbar}
    \int_{-\infty}^{\infty} dp\,
    \exp\!\left[
        \frac{i}{\hbar} p\,(x-x')
        -
        \frac{i}{\hbar} \frac{p^2\,t}{2 m}
    \right]\,,
\label{eq:app_K0_momentum}
\end{equation}
and carrying out the \(x\)- and \(x'\)-integrations explicitly, we
obtain
\begin{equation}
    a_b^{(0)}(t)
    =
    \frac{2 \kappa^3}{\pi \hbar}\,
    \exp\!\left[-\frac{i}{\hbar} s_c(t)\right]
    \int_{-\infty}^{\infty}dp\,
    \frac{
        \exp\!\left[
            -\frac{i p^2 t}{2 m \hbar}
            -\frac{i p\,x_c(t)}{\hbar}
        \right]
    }
    {
        \left[
            \kappa^2 + \left(p/\hbar\right)^2
        \right]
        \left[
            \kappa^2 +
            \left[p + p_c(t)\right]^2/\hbar^2
        \right]
    }\,.
\label{eq:app_ab0_momentum}
\end{equation}

Using \(\eta\) defined in Eq.~\eqref{eq:eta0}, we introduce
\[
    \rho \equiv \frac{p}{\hbar} + \frac{\eta}{2}\,,
\]
whereupon the contact-free amplitude assumes the symmetric form
\begin{equation}
    a_b^{(0)}(t)
    =
    \frac{2 \kappa^3}{\pi}
    \exp\!\left(
        -\frac{i F_0^2\,t^3}{24 m \hbar}
    \right)
    \int_{-\infty}^{\infty} d\rho\,
    \frac{
        \exp\!\left[
            -\frac{i \hbar t}{2 m} \rho^2
        \right]
    }
    {
        \left[
            \left(\rho - \frac{\eta}{2}\right)^2 + \kappa^2
        \right]
        \left[
            \left(\rho + \frac{\eta}{2}\right)^2 + \kappa^2
        \right]
    }\,.
    \label{eq:app_ab0_symmetric}
\end{equation}
The rational factor in Eq.~\eqref{eq:app_ab0_symmetric} has the four
simple poles given in Eq.~\eqref{eq:ab0_poles}. For \(\eta\neq0\),
its partial-fraction decomposition is accordingly
\begin{equation}
    \frac{1}{
        \left[
            \left(\rho - \frac{\eta}{2}\right)^2 + \kappa^2
        \right]
        \left[
            \left(\rho + \frac{\eta}{2}\right)^2 + \kappa^2
        \right]
    }
    =
    \sum_{j=1}^{4}
    \frac{c_j}{\rho - z_j}\,,
\label{eq:app_partial_fraction}
\end{equation}
where the coefficients \(c_j\) are given by
Eq.~\eqref{eq:cj_definition}. Consequently,
\begin{equation}
    a_b^{(0)}(t)
    =
    \frac{2 \kappa^3}{\pi}
    \exp\!\left(
        -\frac{i F_0^2\,t^3}{24 m \hbar}
    \right)
    \sum_{j=1}^{4}
    c_j\,J(z_j;t)\,,
\label{eq:app_ab0_J}
\end{equation}
where
\begin{equation}
    J(z;t)
    \equiv
    \int_{-\infty}^{\infty}
    \frac{
        \exp\!\left[
            -\frac{i \hbar t}{2 m} \rho^2
        \right]
    }{\rho - z}\,d\rho\,.
\label{eq:app_J_definition}
\end{equation}

To evaluate Eq.~\eqref{eq:app_J_definition}, we set
\[
    \alpha_t \equiv \frac{\hbar t}{2 m}\;\;\; ;\;\;\;
    Z \equiv e^{i \pi/4} \sqrt{\alpha_t}\,z
\]
and perform the rotation
\begin{equation}\label{eq:app_Fresnel_rotation}
    \zeta = e^{i \pi/4} \sqrt{\alpha_t}\,\rho\,.
\end{equation}
Since
\[
    d\rho =
    \frac{e^{-i \pi/4}}{\sqrt{\alpha_t}}\,d\zeta
    \;\;\; ;\;\;\;
    \rho - z =
    \frac{e^{-i \pi/4}}{\sqrt{\alpha_t}}(\zeta - Z)\,,
\]
it follows directly that
\[
    \frac{d\rho}{\rho - z}
    =
    \frac{d\zeta}{\zeta - Z}\,.
\]
Thus, the Jacobian factor cancels exactly against the corresponding
factor in the transformed denominator, leaving no residual factor
proportional to \(1/\sqrt{\alpha_t}\). The resulting Gaussian
Cauchy-type integral can then be related to the standard real-axis
representation of the Faddeeva function \(w(z)\) defined in
Eq.~\eqref{eq:Faddeeva_definition} \cite{DLM26},
\begin{equation}
    w(Z)
    =
    \frac{1}{\pi i}
    \int_{-\infty}^{\infty}
    \frac{e^{-y^2}}{y-Z}\,dy\,,
    \qquad
    \operatorname{Im}Z>0\,.
\label{eq:app_Faddeeva_integral}
\end{equation}
Because this representation is directly valid for
\(\operatorname{Im} Z > 0\), the corresponding pole contribution
must be retained when continuing to the lower-half-plane case. For
the four poles in Eq.~\eqref{eq:ab0_poles}, this is encoded by the
pole prescription in Eq.~\eqref{eq:Zj_sigmaj}, giving
\begin{equation}
    J(z_j;t)
    =
    i\pi\,\sigma_j\,
    w\!\left(\sigma_j Z_j\right)\,.
\label{eq:app_J_Faddeeva}
\end{equation}
Therefore, \(\sigma_j\) accounts for whether the original pole
\(z_j\) lies in the upper or lower half-plane and ensures the
appropriate continuation of the Faddeeva representation.

Substituting Eq.~\eqref{eq:app_J_Faddeeva} into
Eq.~\eqref{eq:app_ab0_J} finally recovers
Eq.~\eqref{eq:ab0_Faddeeva}. At \(\eta = 0\), however, the pairs of
simple poles coalesce. Hence, the individual coefficients \(c_j\)
should not be evaluated by setting \(\eta = 0\) directly.
Eq.~\eqref{eq:ab0_Faddeeva} is instead understood through the
continuous limit \(\eta \rightarrow 0\).

\section{Field-Free Consistency Check and Exact Pole Prescription}
\label{app:field_free_consistency}
%
The field-free limit provides a stringent consistency check of the
complete finite-time bound-state amplitude. Since the attractive
delta-function Hamiltonian has a single bound state with energy
given in Eq.~\eqref{eq:bound_state_energy}, the exact field-free
evolution must satisfy \(a_b(t) = e^{i \Omega t}\) and
\(\left|a_b(t)\right|^2 = 1\), where \(\Omega \equiv \hbar
\kappa^2/(2 m)\). We verify this result directly from the contact
and contact-free contributions.

For \(F_0 = 0\), the projection block entering
Eq.~\eqref{eq:ab_delta_single_integral0} follows from
Eqs.~\eqref{eq:X-final} and \eqref{eq:I_sigma_q0} as
\begin{equation}
    X\!\left(u,t-\frac{u}{2}\right)
    =
    \sqrt{\kappa}\,
    e^{i \Omega u}\,
    \operatorname{erfc}\!\left(\sqrt{i \Omega u}\right)\,.
\end{equation}
Likewise, Eq.~\eqref{eq:q_0-1} gives the zeroth-order contact
amplitude
\begin{equation}
    q_0(\xi)
    =
    \sqrt{\kappa}\,
    e^{i \Omega \xi}\,
    \operatorname{erfc}\!\left(\sqrt{i \Omega \xi}\right)\,.
\end{equation}
The quantity entering Eq.~\eqref{eq:ab_delta_single_integral0},
however, is the full physical contact amplitude \(q(\xi)\), rather
than its zeroth-order part \(q_0(\xi)\). In the field-free limit,
\(\Theta(\xi) = 0\), \(Q(\xi) = q(\xi)\), and \(Q_0(\xi) =
q_0(\xi)\), while the field-dependent resolvent reduces as
\(R_{\scriptstyle F} \to R_0\). The exact resolvent representation
in Eq.~\eqref{eq:q_resolvent} therefore gives
\begin{equation}
    q(\xi)
    =
    q_0(\xi)
    +
    \int_0^\xi dv\,
    R_0(v)\, q_0(\xi - v)
    =
    \sqrt{\kappa}\,e^{i \Omega \xi}\,.
\label{eq:q_field_free_resolvent}
\end{equation}
This agrees with the exact field-free contact-point evolution
obtained independently in Appendix~\ref{app:field_free_case}. The
contact contribution therefore reduces to
\begin{equation}
    a_b^{(\delta)}(t)
    =
    2 i \Omega\,e^{i \Omega t}
    \int_0^t du\,
    \operatorname{erfc}\!\left(\sqrt{i \Omega u}\right)
    =
    e^{i \Omega t}
    \left[
        1 + (2 i \Omega t - 1)
        \operatorname{erfc}\!\left(\sqrt{i \Omega t}\right)
    \right]
    -
    \sqrt{\frac{4 i \Omega t}{\pi}}\,.
    \label{eq:ab_delta_field_free_integral}
\end{equation}

We next evaluate the contact-free contribution and, at the same
time, verify that the exact \(\sigma_j\)-prescription used in
Eq.~\eqref{eq:ab0_Faddeeva} reproduces the correct field-free limit.
Setting \(F_0 = 0\) in Eq.~\eqref{eq:app_ab0_symmetric} gives
\begin{equation}
    a_b^{(0)}(t)
    =
    \frac{2 \kappa^3}{\pi}
    \int_{-\infty}^{\infty}
    \frac{e^{-i \alpha_t \rho^2}}
         {(\rho^2 + \kappa^2)^2}\,d\rho\,,
    \qquad
    \alpha_t = \frac{\hbar t}{2 m}\,.
    \label{eq:ab0_symmetric_field_free}
\end{equation}
Rather than substituting \(\eta = 0\) directly into the four-pole
partial-fraction form, we evaluate this symmetric field-free
integral directly. Define
\[
    I_1(\kappa)
    =
    \int_{-\infty}^{\infty}
    \frac{e^{-i \alpha_t \rho^2}}
         {\rho^2 + \kappa^2}\,d\rho\,.
\]
Using
\[
    \frac{1}{\rho^2 + \kappa^2}
    =
    \frac{1}{2 i \kappa}
    \left(
        \frac{1}{\rho - i \kappa}
        -
        \frac{1}{\rho + i \kappa}
    \right)\,,
\]
the prescription \(J(z;t) = i \pi \sigma\,w(\sigma Z)\) in
Eq.~\eqref{eq:app_J_Faddeeva}, with \(\sigma =
\operatorname{sgn}(\operatorname{Im} z)\), gives
\[
    J(i\kappa;t)=i\pi\,w(W)
    \;\;\; ;\;\;\;
    J(-i\kappa;t)=-i\pi\,w(W)\,,
\]
where \(W = e^{3 i \pi/4}\,\kappa \sqrt{\alpha_t}\). Thus, both
half-plane contributions are incorporated with their proper signs,
and
\[
    I_1(\kappa)
    =
    \frac{\pi}{\kappa}\,w(W)\,.
\]
Since
\[
    W^2 = -i \alpha_t \kappa^2\,,
    \qquad
    -i W = \sqrt{i \Omega t}\,,
    \qquad
    \alpha_t \kappa^2 = \Omega t\,,
\]
we obtain via Eq.~\eqref{eq:Faddeeva_definition}
\[
    I_1(\kappa)
    =
    \frac{\pi}{\kappa}\,
    e^{i \Omega t}\,
    \operatorname{erfc}\!\left(\sqrt{i \Omega t}\right)\,.
\]
Using
\[
    \frac{1}{(\rho^2 + \kappa^2)^2}
    =
    -\frac{1}{2 \kappa}
    \frac{\partial}{\partial\kappa}
    \frac{1}{\rho^2 + \kappa^2}\,,
\]
Eq.~\eqref{eq:ab0_symmetric_field_free} becomes
\begin{equation}
    a_b^{(0)}(t)
    =
    -\frac{\kappa^2}{\pi}
    \frac{\partial I_1(\kappa)}{\partial\kappa}
    =
    (1 - 2 i \Omega t)
    e^{i \Omega t}
    \operatorname{erfc}\!\left(\sqrt{i \Omega t}\right)
    +
    \sqrt{\frac{4 i \Omega t}{\pi}}\,.
    \label{eq:ab0_field_free}
\end{equation}

Combining Eqs.~\eqref{eq:ab_delta_field_free_integral} and
\eqref{eq:ab0_field_free}, the \(\operatorname{erfc}\)-dependent and
square-root terms cancel exactly, leaving
\begin{equation}
    a_b(t)
    =
    a_b^{(0)}(t) + a_b^{(\delta)}(t)
    =
    e^{i \Omega t}\,,
    \qquad
    \left|a_b(t)\right|^2 = 1\,.
    \label{eq:field_free_complete_amplitude}
\end{equation}
Thus, the exact field-free bound-state evolution is recovered from
the independently calculated contact and contact-free contributions,
while the \(\sigma_j\)-prescription is confirmed to incorporate the
lower-half-plane contribution correctly in the Faddeeva
representation. This provides an independent consistency check of
the complete finite-time amplitude.
\bibliographystyle{aipnum4-2}
\bibliography{literature}
\end{document}